*Full version-2 for ArXiv*

# Correlation between corrugation-induced flexoelectric polarization and conductivity of low-dimensional transition metal dichalcogenides

Anna N. Morozovska[1*], Eugene A. Eliseev[2], Hanna V. Shevliakova[1,3], Yaroslava Yu. Lopatina[1], Galina I. Dovbeshko[1], Maya D. Glinchuk[2], Yunseok Kim[4†], and Sergei V. Kalinin[5,‡]

[1] *Institute of Physics, National Academy of Sciences of Ukraine,*
*46, pr. Nauky, 03028 Kyiv, Ukraine*

[2] *Institute for Problems of Materials Science, National Academy of Sciences of Ukraine,*
*Krjijanovskogo 3, 03142 Kyiv, Ukraine*

[3] *Department of Microelectronics, National Technical University of Ukraine "Igor Sikorsky Kyiv Polytechnic Institute", Kyiv, Ukraine*

[4] *School of Advanced Materials Science and Engineering, Sungkyunkwan University (SKKU), Suwon 16419, Republic of Korea*

[5] *The Center for Nanophase Materials Sciences, Oak Ridge National Laboratory,*
*Oak Ridge, TN 37922*

---

[*] Corresponding author 1: anna.n.morozovska@gmail.com

[†] Corresponding author 2: yunseokkim@skku.edu

[‡] Corresponding author 3: sergei2@ornl.gov

**Abstract**

Tunability of polar and semiconducting properties of low-dimensional transition metal dichalcogenides (TMDs) have propelled them to the forefront of fundamental and applied physical research. These materials can vary from non-polar to ferroelectric, and from direct-band semiconductor to metallic. However, in addition to classical controls such as composition, doping, and field effect in TMDs the additional degrees of freedom emerge due to the curvature-induced electron redistribution and associated changes in electronic properties. Here we numerically explore the elastic and electric fields, flexoelectric polarization and free charge density for a TMD nanoflake placed on a rough substrate with a sinusoidal profile of the corrugation using finite element modelling (FEM). Numerical results for different flake thickness and corrugation depth yield insight into the flexoelectric nature of the out-of-plane electric polarization and establish the unambiguous correlation between the polarization and static conductivity modulation caused by inhomogeneous elastic strains coupled with deformation potential and strain gradients, which evolve in TMD nanoflake due to the adhesion between the flake surface and corrugated substrate. We revealed a pronounced maximum at the thickness dependences of the electron and hole conductivity of $MoS_2$ and $MoTe_2$ nanoflakes placed on a corrugated substrate, which opens the way for their geometry optimization towards significant improvement their polar and electronic properties, necessary for their advanced applications in nanoelectronics and memory devices. Specifically, obtained results can be useful for elaboration of nanoscale straintronic devices based on the bended $MoS_2$, $MoTe_2$ and MoSTe nanoflakes, such as diodes and bipolar transistors with a bending-controllable sharpness of p-n junctions.

# I. INTRODUCTION

Tunability of polar and semiconducting properties of low-dimensional (**LD**) transition metal dichalcogenides (**TMDs**) with a chemical formula $MX_2$ (M – metal Mo, W, Re; X – chalcogen S, Se, Te) and Janus-compounds (**JC**) with a chemical formula MXY (X, Y – chalcogens) in the form of monolayers and nanoflakes have propelled them to the forefront of fundamental and applied physical research [1, 2, 3, 4]. These materials offer a broad gamut of properties varying from non-polar to ferroelectric, and from direct-band semiconductor to metallic. Remarkedly that LD semiconductor materials, such as graphene, $MX_2$ and MXY, are ideal candidates for the strain engineering [5] and straintronics [6], because that the strain-induced conductive domain walls here can act as mobile charged channels similarly to the "domain wall nanoelectronics" in multiferroic thin films [7, 8, 9] and graphene-on-ferroelectric nanostructures [10, 11].

Layered TMDs in the form of bulk materials are typically non-polar centrosymmetric semiconductors with a relatively wide band gap ~(1.1 – 2) eV [1]. However, on transition from the bulk to the nanoscale additional orderings emerge, which can be non-polar or piezoelectric, or even ferroelectric [12, 13, 14], semiconductive, semi-metallic or metallic [15, 16, 17], as are found in different structural phases (polymorphs) of TMD monolayers. Similarly, in addition to classical controlling factors, such as composition, doping, and field effect, the additional degrees of freedom emerge in TMDs due to the curvature-induced electron redistribution and associated changes in electronic properties (see e.g. [18, 19]). There are very prospective theoretical and experimental possibilities for tuning the structural, polar and electronic properties of LD-TMDs by application of homogeneous [20, 21] and inhomogeneous [22, 23] elastic strains as well as by the doping effect [24] on the structural phases and electronic properties of the LD-TMDs. However, these possibilities are almost not systematized and mostly empirical.

From theoretical perspective, a number of first-principle studies explored surface-induced piezoelectricity [25, 26] and out-of-plane ferroelectric polarization [14, 27] in several $MX_2$. As predicted by several authors [19, 28], and later discovered experimentally [22], an elastic bending can induce polar phenomena in LD-TMDs, either intrinsic or induced by the external structural and charge disorder. Large curvatures enabled by small bending stiffnesses can give rise to a significant polarization induced by a flexoelectric effect [29, 30]. The bending-induced out-of-plane dipole moment with density $p \sim (0.01 - 0.4)$ C/nm and flexoelectric polarization $P \sim (0.1 - 2) \mu C/cm^2$ were calculated from the first principles for $MoS_2$ [14], $WS_2$ [30], and $WTe_2$ [31, 32], respectively. Therefore, it is quite possible that the strongly inhomogeneous spontaneous deformation of LD-TMD, which causes the appearance of their spontaneous polarization, is due to the flexoelectric coupling [1, 22, 34]. Let us focus on several recent examples.

Previously, we describe analytically MX$_2$ structural phases and discuss the mechanisms of ferroelectric state appearance using Landau-Ginzburg-Devonshire-type (**LGD-type**) approach [33]. The LGD-type thermodynamic analysis suggests that out-of-plane ferroelectricity can exist in many phases of LD-TMDs with the switchable polarization being proportional to the out-of-plane order parameter, and further predict that the domain walls in LD-TDMs should become conductive above a certain strain threshold. Despite the progress, the nature of one-dimensional piezoelectricity and ferroelectricity has not been elucidated, and the mesoscopic analytical theory of polar and electronic phenomena in LD-TMD has not been constructed. Such a theory is necessary to control and predict the physical properties of LD-TMD and JC for their novel applications in nanoelectronics and advanced memories.

From a crystallographic and symmetry point of view, only the in-plane piezoelectricity can exist in a geometrically flat centrosymmetric MX$_2$. Even though symmetry, the tunable out-of-plane piezoelectricity induced by the local flexoelectricity was observed by Kang et al. [22] in semiconducting 2H-MoTe$_2$ flakes by creating surface corrugation. Surface corrugation-mediated flexoelectricity tuning can be applied to other two-dimensional or thin-layered materials and, furthermore, the results could provide useful information on the interweaving nature between mechanical stimulus and electric dipole in low-dimensional materials. Next, Kang et al. [1] demonstrated the creation of strong out-of-plane piezoelectricity in semiconducting 2H-MoTe$_2$ flakes by an artificial atomic-scale symmetry breaking realized through Te vacancy formation, and confirmed by DFT calculations.

Recently [34] we developed LGD-type theory for the description of polar phenomena in LD-TMDs, specifically exploring flexoelectric origin of the polarization induced by a spontaneous bending and by inversion symmetry breaking due to the interactions with substrate. We consider the appearance of the spontaneous out-of-plane polarization due to the flexoelectric coupling with the strain gradient of the spontaneous surface rippling and surface-induced piezoelectricity. Performed calculations proved that the out-of-plane spontaneous polarization, originated from flexoelectric effect in a rippled TMD, is bistable and reversible by a non-uniform electric field.

This work is devoted to the establishment of correlation between polar and electronic properties of LD-TMDs and JCs, which is almost unexplored. Using finite element modeling (**FEM**) we calculate the elastic and electric fields, flexoelectric polarization and free charge density for a TMD (or JC) nanoflake placed on a rough substrate with a sinusoidal profile of the corrugation. Analysis of FEM results obtained for the different flake thickness (varying from 10 layers to 300 nm in accordance with experiment [22]) and corrugation depth (varying from 0 to 50 nm according to typical technological conditions [35]), allows to corroborate the flexoelectric nature of the out-of-plane electric polarization and establish the unambiguous correlation between the polarization and static conductivity modulation caused by inhomogeneous elastic strains coupled with deformation potential, and strain gradients, which evolve in the nanoflake due to the adhesion between the flake surface and corrugated substrate.

## II. THEORETICAL MODEL AND MATERIAL PARAMETERS

Here we explore the effect of bending-induced flexoelectric polarization of $MX_2$ nanoflakes originated from the adhesion to a corrugated substrate. For JCs, the flexoelectric polarization can be superposed with the piezoelectric polarization coming from inversion symmetry breaking at the MXY surface. The overall picture of the bending-induced flexoelectric polarization and surface-induced piezoelectric polarization in and $MX_2$ and MXY is shown schematically in **Fig. 1**.

A flat single-layer (**SL**) with a chemical formula $MX_2$, where M is a transition metal, X is a chalcogenide is shown in **Fig. 1a**. A projection of M ions is located at the middle line between X ions projection. The inhomogeneous strain is absent in a flat SL $MX_2$. Since the effective (Born or Bader) charges $Q$ of M and X ions are opposite, namely $Q_M = -2Q_X$, the total polarization is absent. For the case of a stretched flat Janus-compound with a chemical formula XMY, the top X and bottom Y chalcogen ions are in non-equivalent conditions despite their planes are parallel (see **Fig. 1d**). While the total charge is zero, $Q_M = -Q_X - Q_Y$, the surface-induced piezoelectric effect can induce surface dipoles resulting into the nonzero total out-of-plane polarization $P$ shown by a dashed black curve and arrows in **Fig. 1d**.

A mechanically free SL $MX_2$ and MXY, where all X, Y and M ions positions can spontaneously reconstruct to the rippled state with the lowest energy, are shown in **Fig. 1b** and **1f**, respectively. Since the force matrix is different for the "light" X (or Y) and "heavy" M ions, the amplitudes of X (or Y) and M displacements are different for the periodic ripples. Since the total effective charge is absent the periodic displacement of X (or Y) and M ions induces the out-of-plane polarization modulation (dashed black curve and arrows) only due to the flexoelectric coupling.

The flexoelectric coupling exists for all possible symmetries [41] and geometries of the TMD and JC; hence the effect always contributes to the total polarization, as shown in **Fig. 1b-1d** and **1f-1h**.

The surface-induced piezoelectric effect contributes to the total polarization of a SL-MX$_2$ only for the case on non-equivalent conditions for the top and bottom chalcogen layers, that is the case of one surface free and other clamped to a flat rigid substrate (see **Fig. 1c**). The surface-induced out-of-plane piezoelectric effect always contribute to the total polarization of a SL MXY, due to the different elastic force constants for the top "Y" and bottom "X" chalcogen layers (see **Fig. 1e**-**1g**).

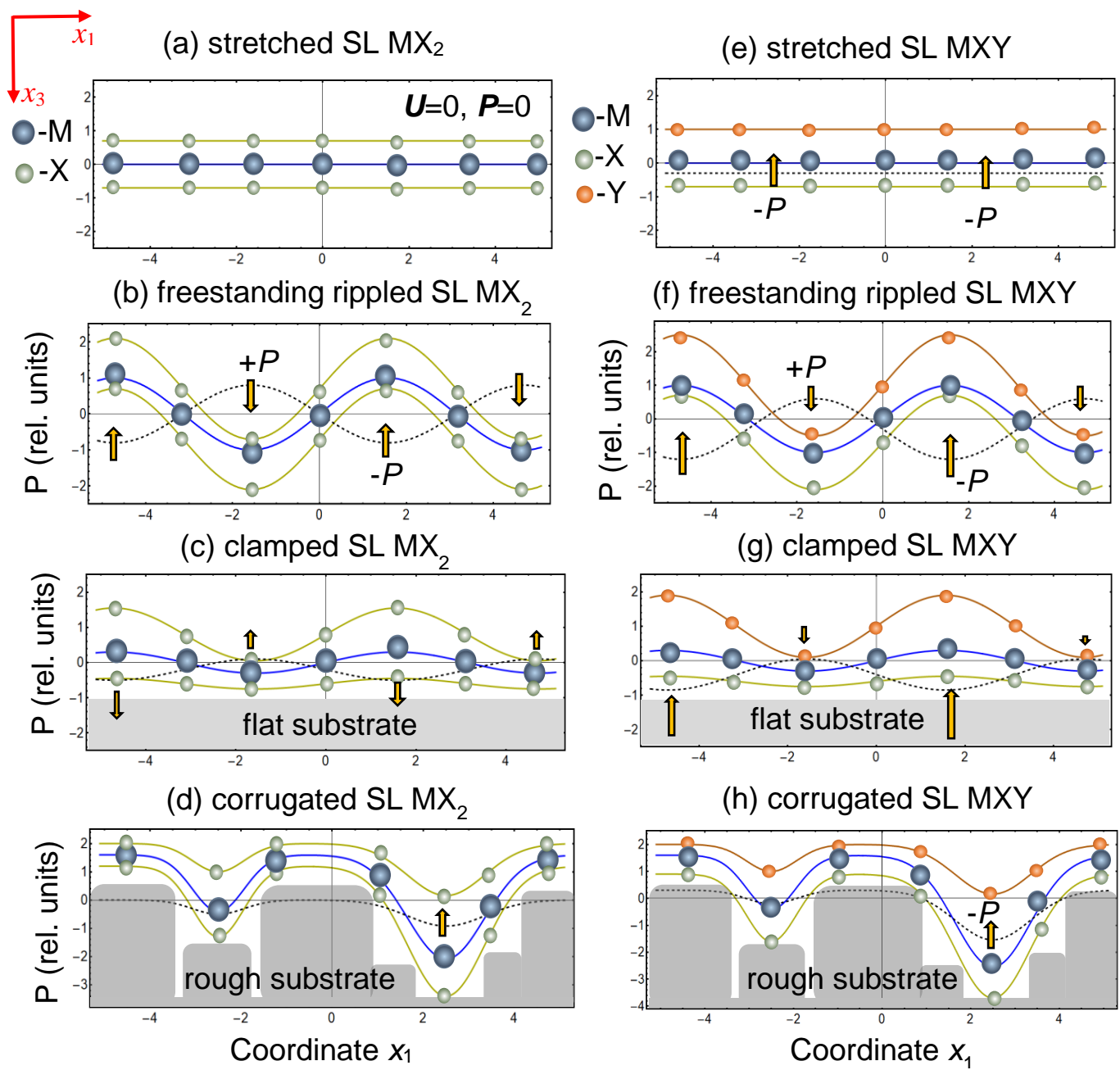

**FIGURE 1. Bending-induced out-of-plane polarization in a SL MX₂ (a-d) and Janus compound YMX (e-h).** M is a transition metal, X and Y are chalcogen atoms. M ions are positively charged, X and Y ions are negatively charged. Orange arrows indicate the direction of emerging out-of-plane dipole polarization P, which profile is shown by a dashed black curve. A relative atomic displacement *U* and polarization *P* are absent for a flat stretched SL MX₂ **(a)**; and a small uniform *P* can appear for YMX **(e)**. A freestanding SL MX₂ **(b)** and MXY **(f)**. **(c)** Rippled and strained SL MX₂ **(c, d)** and MXY **(g, h)** clamped to a flat **(c,g)** or rough **(d,h)** rigid substrate. Parts **(a)-(c)** are reprinted from Ref. [34].

Since the bottom X ions are bonded to the substrate [36] by different types of van der Waals (**vdW**) forces [37, 38], in the case of strong adhesion [39, 40] they remained almost "clamped" to the atomic planes of a flat (i.e. atomically smooth) substrate or to the islands of atoms of a rough (i.e. corrugated) substrate. For the case of rough substrates used in e.g. experiments [1, 22] the inhomogeneous strain can exist in the clamped sections, while the suspended sections can relax freely (see **Fig. 1d** and **1h**). The amplitude of the clamped atoms vertical displacement is very small there, a more visible change of position is possible for the middle and top ions. A significant reconstruction occurs for the ions at "suspended" sections, where the different amplitudes of the displacement of bottom X, middle M ions and top X (or Y) ions induce the polarization (dashed black curve and arrows) due to the flexoelectric coupling in MX₂, and both flexoelectric coupling and surface-induced piezoelectric effect in MXY.

The out-of-plane component $P_3$ of a SL TMD or JC shown in **Figs. 1b-h** can be estimated by a continuum media approach [34], which is applicable even for ultrathin layers [22, 21], and becomes quantitative for a nanoflake with the thickness more than 10 layers. Within this framework [34], $P_3$ comprises the flexoelectric and piezoelectric contributions. The first one is proportional to the second derivatives of elastic displacement in space and time multiplied by static and dynamic flexoelectric coefficients, respectively. The second one is proportional to surface-induced piezoelectric coefficient multiplied by an elastic strain. So:

$$P_3(\vec{x}, t) \cong f_{3jkl} \frac{\partial^2 U_j(\vec{x},t)}{\partial x_k \partial x_l} - \mu_{3j} \frac{\partial^2 U_j(\vec{x},t)}{\partial t^2} + \frac{e^S_{3jk}}{t} \frac{\partial U_j(\vec{x},t)}{\partial x_k}. \qquad (1)$$

Here $U_j(\vec{x})$ is the mechanical displacement of the flake, which includes a spontaneous and misfit contributions, $f_{ijkl}$ is the static flexoelectric tensor [41] determined by the microscopic properties of the material [42, 43], $\mu_{ij}$ is the dynamic flexoelectric tensor [44], $e^S_{ijk}$ is the tensor of the surface-induced piezoelectric effect [45, 46], measured in pC/m and $t$ is the flake thickness [47]. It follows from the symmetry considerations, $e^s_{3jk}$ is zero for a flat centrosymmetric $MX_2$ with inversion axis "3", it was estimated as several −2.4 pC/m for a reconstructed SL-$MoS_2$ in 1T' phase [14]. It is much higher ($e^s_{31}$~50 pC/m) for an asymmetric Janus compound MoSTe [27, 31]. Both these values are much smaller than in-plane components, $e^s_{1jk}$~(300 − 500) pC/m, estimated for $MoS_2$, $MoTe_2$, and MoSTe [14, 27].

The linear partial differential equation relating the mechanical displacement $U_j(\vec{x})$ of the flake points and its elastic stress $\sigma_{ij}$ has the form:

$$\rho \frac{\partial^2 U_i}{\partial t^2} + \frac{\partial}{\partial x_j} \sigma_{ij}(\vec{x},t) = \mu_{ij} \frac{\partial^2 P_j}{\partial t^2} + \frac{\partial}{\partial x_i} \psi(\vec{x},t), \qquad (2)$$

where $\rho$ is the density of material and $\psi(\vec{x},t)$ is an elastic force density (with dimensionality N/m$^3$) in the nanoflake. The physical origin of these forces can include the atomic reconstruction of the free surface, as well as adhesion (e.g. vdW force) [37 – 40] to the flat or corrugated substrate. Within a continuum media approach, we used here, the force cannot be derived, but rather postulated.

For a static case considered here, Eq.(1) reduces to

$$P_3(\vec{x}) \cong f_{3jkl} \frac{\partial}{\partial x_j} u_{kl}(\vec{x}) + \frac{e^s_{3jk}}{t} u_{jk}(\vec{x}), \qquad (3a)$$

where $u_{kl} = \frac{1}{2}\left(\frac{\partial U_k}{\partial x_l} + \frac{\partial U_l}{\partial x_k}\right)$ is an elastic strain tensor. The coupling between the stress $\sigma_{ij}$ and strain $u_{kl}$ is given by the generalized Hooke's law [48],

$$\sigma_{ij} = c_{ijkl} u_{kl} + F_{ijmn} \frac{\partial P_m}{\partial x_n} + h^s_{3ij} P_3 \approx c_{ijkl} u_{kl} + O\left(f^2_{3ij}, (h^s_{3ij})^2\right) \qquad (3b)$$

where $c_{ijkl}$ is the tensor of elastic stiffness, $F_{ijmn}$ is a flexoelectric tensor (in V), and $h^s_{mij}$ is a surface-induced piezoelectric stress tensor (in V), which is absent for a centrosymmetric $MX_2$ and present for Janus MXY. The approximate equality is the decoupling approximation, valid for small flexoelectric and surface piezoelectric effects, which are indeed small in the considered case.

Using the decoupling approximation in Eq.(3b), Eq.(2) yields Lame equation:

$$c_{ijkl} \frac{\partial^2 U_k(\vec{x})}{\partial x_j \partial x_l} = \frac{\partial}{\partial x_i} \psi(\vec{x} - \vec{x}_s), \qquad (3c)$$

where $\vec{x}_s$ is the corrugation profile of a rough substrate (see **Fig. 2**).

To solve Eq. (3c), we need to define the adhesion force $\psi(\vec{x})$ between the layers of X, M or/and Y atoms, and an atomically flat or corrugated substrate. Here we use the assumption that the constitutive parts of vdW forces [37] corresponding to dipole-dipole, and dispersive-dipole interactions and London dispersive force, determine the atomic repulsion at ultra-small distances, and their attraction at higher distances. In this approximation, the adhesion force (per unit volume, in J/m$^3$) between the flake and substrate can be approximated by power law attraction with a prefactor – contact adhesive stiffness $S_A$ at distances $|\vec{x} - \vec{x}_s| > \delta h_0$, and strong repulsion at distances $|\vec{x} - \vec{x}_s| < \delta h_0$ (see also **Appendix A** in **Suppl. Mat.** [49]).

Using FEM, we analyze the scenario where TMD nanoflakes are placed on a rough substrate with a sinusoidal corrugation profile $\vec{x}_s = \left\{ A \cos\left(\frac{2\pi}{\lambda} x_1\right), 0, 0 \right\}$. For sufficiently strong adhesion forces, the flake displacement, strain, strain gradient, and out-of-plane electric polarization induced by the flexoelectric coupling are modulated by the corrugation profile, being maximal at the corrugation peaks and minimal in the valleys (see **Fig. 2**). Material parameters of several $MX_2$, MXY, and substrate, which are used for FEM, are listed in **Table I**. All calculations were performed at room temperature $T = 300$ K.

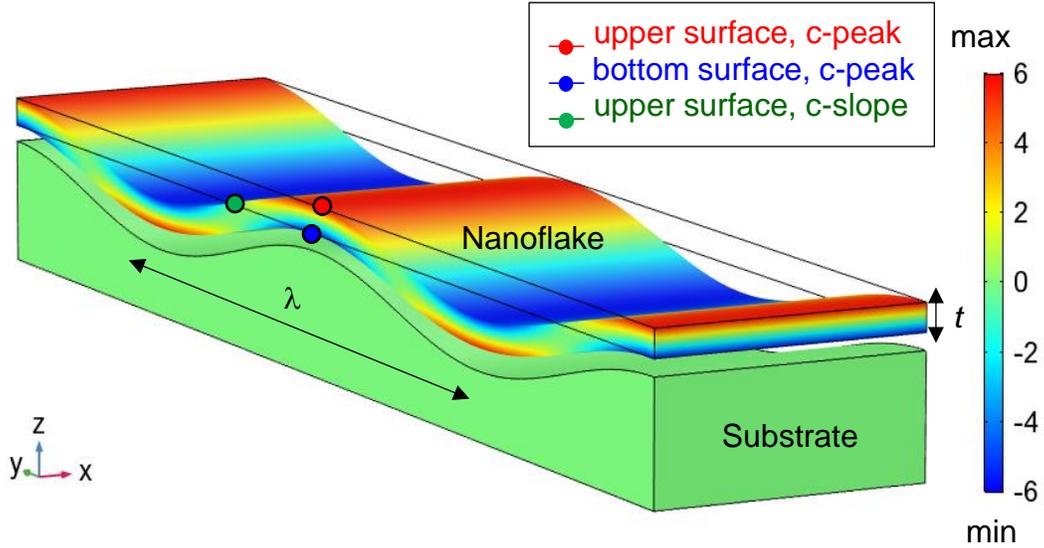

**FIGURE 2. Considered geometry. A nanoflake of TMD is placed on a corrugated substrate.** Right scale schematically shows the variation of electric polarization, induced by the flexoelectric effect, and electric conductivity, induced by inhomogeneous elastic strain via the deformation potential. The red and blue circles located at the upper and bottom surfaces of the nanoflake, which is placed at the peak of the substrate corrugation, are abbreviated as "**c-peak**". The green circles located at the corrugation slope is abbreviated as "**c-slope**". Further we analyze the behavior of the physical properties in these three characteristic points at the nanoflake surface.

The inhomogeneous elastic strain should induce the changes of the electron and hole charge density, $n(\vec{x})$ and $p(\vec{x})$, in the flake due to e.g. the coupling with band structure via a deformation potential:

$$\frac{n(\vec{x})}{n_0} = \frac{1}{n_0}\int_0^\infty \frac{g_n(\varepsilon)d\varepsilon}{1+exp\left(\frac{\varepsilon-E_F+E_C+\Sigma_{ij}^e u_{ij}(\vec{x})-e\varphi}{k_B T}\right)} \sim exp\left(-\frac{\Sigma_{ij}^e u_{ij}(\vec{x})-e\varphi}{k_B T}\right), \quad (4a)$$

$$\frac{p(\vec{x})}{p_0} = \frac{1}{p_0}\int_0^\infty \frac{g_p(\varepsilon)d\varepsilon}{1+exp\left(\frac{-\varepsilon+E_F-E_V+\Sigma_{ij}^h u_{ij}(\vec{x})+e\varphi}{k_B T}\right)} \sim exp\left(-\frac{\Sigma_{ij}^h u_{ij}(\vec{x})+e\varphi}{k_B T}\right), \quad (4b)$$

where $n_0 = \int_0^\infty \frac{g_n(\varepsilon)d\varepsilon}{1+exp\left(\frac{\varepsilon-E_F+E_C}{k_B T}\right)}$ and $p_0 = \int_0^\infty \frac{g_p(\varepsilon)d\varepsilon}{1+exp\left(\frac{-\varepsilon+E_F-E_V}{k_B T}\right)}$ are the electron and hole concentrations in the unstrained nanomaterial, $g_n(\varepsilon)$ and $g_p(\varepsilon)$ are the densities of electron and hole states, $E_F$ is the Fermi energy, $k_B = 1.3807 \times 10^{-23}$ J/K is a Boltzmann constant, $\Sigma_{ij}^e$ and $\Sigma_{ij}^h$ are the deformation potential tensors for electrons and holes, respectively, which values can be estimated from the first principles and listed in **Table I** for several MX$_2$. The proportionality in Eq.(5) is valid for a Boltzmann-Plank-Nernst statistics of non-degenerated carriers. Since the band gap of MX$_2$ and MXY is about (1 – 2) eV, we can expect that rippled MX$_2$ and MXY should become conductive, once the spontaneous strain value exceeds several %, which makes the product $-\Sigma_{ij}^{e,h} u_{ij} \gg k_B T$.

**Table I.** Materials constants for a LD MX$_2$ and MXY, and rough substrate

| Material | LD MoS$_2$ | LD MoTe$_2$ | LD or LS MoSTe |
|---|---|---|---|
| Symmetry/phase * | 6/mmm, 2H or 1H | 6/mmm, 2H or 1H | 3m, 1H |
| Lattice constants (nm) | c=1.1, a=0.32 | c=1.4, a=0.35 | c=1.3, a=b=0.34 |
| Elastic modules (GPa) | c$_{11}$=134.3, c$_{12}$=33.3 | c$_{11}$=139.0, c$_{12}$=71.0 (for v>0) | c$_{11}$=116, c$_{12}$=28 |
| Young module $E$ (GPa) | 250 Ref.[50]<br>136 Ref.[51] | (90 – 110) Ref.[52]<br>we use 100 | 105 |
| Poisson ratio v | −0.07 [53] | 0.34 [52]<br>(but v= −0.08 from [53]) | 0.24 – 0.30 [54] |
| Effective piezo-response $d_{ijk}^{eff}$ (pm/V) | $d_{11}^{eff}$ =5.76 [14]<br>$d_{31}^{eff}$ = −0.018 [14] | $d_{33}^{eff} \approx 3$ [22]<br>$d_{11}^{eff}$=9.13 [55]<br>$d_{11}^{eff}$=7.39 [26] | $d_{11}^{eff}$ =5.1 [54]<br>$d_{31}^{eff}$ =0.4 [54] |
| surface-induced out-of-plane piezoelectric coefficients $e_{ijk}^s$ (pC/m) | $e_{31}^s = 0$<br>$e_{33}^s = 0$ | $e_{31}^s = 0$<br>$e_{33}^s = 0$ | $e_{31}^s$= 50 pC/m<br>$\|e_{33}^s\| < 5$ pC/m<br>[27, 31, 54] |

| Flexoelectric coefficient $f_{ijkl}$ (nC/m) Ref.[56] | $f_{13}$ =4.44<br>$f_{13}$ =4.44 | $f_{13}$ =9.50<br>$f_{11}$ = 9.50 | unknown, set equal to $f_{13} = f_{33}$ =6.47 |
|---|---|---|---|
| Deformation potential $\Sigma_{ij}$ (eV) | $\Sigma_h$= 5.61 [27]<br>$\Sigma_e$= −11.14 [27]<br>$\Sigma_g$= −(13.8 − 17) Ref.[20] | $\Sigma_h$= 4.2 (this work)<br>$\Sigma_e$= −7 (this work)<br>$\Sigma_g$= −(3.8 − 11) Ref.[20] | unknown, set equal to $\Sigma_h$= −5 (this work)<br>$\Sigma_e$= −7 (this work) |
| Substrate | Adhesive stiffness varies in the range $S_A$ =(20 − 2) GPa, adhesion repulsive separation is equal to $\delta h_0 = 1.1$ nm. Materials vary from gold to silica oxide. The average period of corrugation is $\lambda$ =0.5 μm, the amplitude $A$ of substrate roughness varies from 0 to 11 nm. A substrate thickness (200 – 500) nm | | |

**\*** We regard that LD MoX$_2$ and MXY exist in the semiconducting H phases (2H or 1H) under normal conditions.

### III. CORRELATION BETWEEN POLAR AND ELECTRONIC PROPERTIES OF MX$_2$ AND MXY NANOFLAKES

Spatial distributions and profiles of elastic strain, its gradient, polarization and charge density of the TMDs and JCs nanoflakes are shown in **Figs. 3-7**. Additional details are shown in **Figs. S1-S4** in **Appendix B** [49] and briefly described here. One period of the substrate corrugation profile $\vec{x}_s = \left\{A \cos\left(\frac{2\pi}{\lambda} x_1\right), 0, 0\right\}$ is shown in all these figures, and the unambiguous correlations between the maximal/minimal values of the strain, strain gradient, polarization and charge density are clearly seen at the period.

The cross-sections of bending-induced elastic displacement, strain components, out-of-plane polarization induced by the flexoelectric coupling, and carrier density distributions are shown in **Figs. 3a-3f** for MoS$_2$ and **Figs. 3g-3l** MoTe$_2$ nanoflakes of the same thickness (~28 nm) on the rough Au-substrate, respectively. It is seen that the nanoflake displacement, strain, strain gradient, out-of-plane electric polarization and charge density are modulated by the corrugation profile, being maximal at the corrugation peaks, zero at the slopes, and minimal in the valleys. At that the relative increase of the electron density for MoS$_2$ (more than 10 times) is much higher than that for a MoTe$_2$ (less than 5 times), compare **Fig. 3f** and **3l**. The regions enriched by holes (more than 10 times) exists in MoS$_2$ nanoflakes. Hence the appearance of the p-n junctions between the n-type and p-type regions are possible in MX$_2$ nanoflakes, similarly to the case of graphene-on-ferroelectric [10, 11]. The distinct feature of MX$_2$ nanoflakes-on-corrugated substrate is that the width, diffuseness or sharpness of the p-n junctions can be controlled by the flake thickness $t$, and by the relative substrate corrugation, $A/\lambda$.

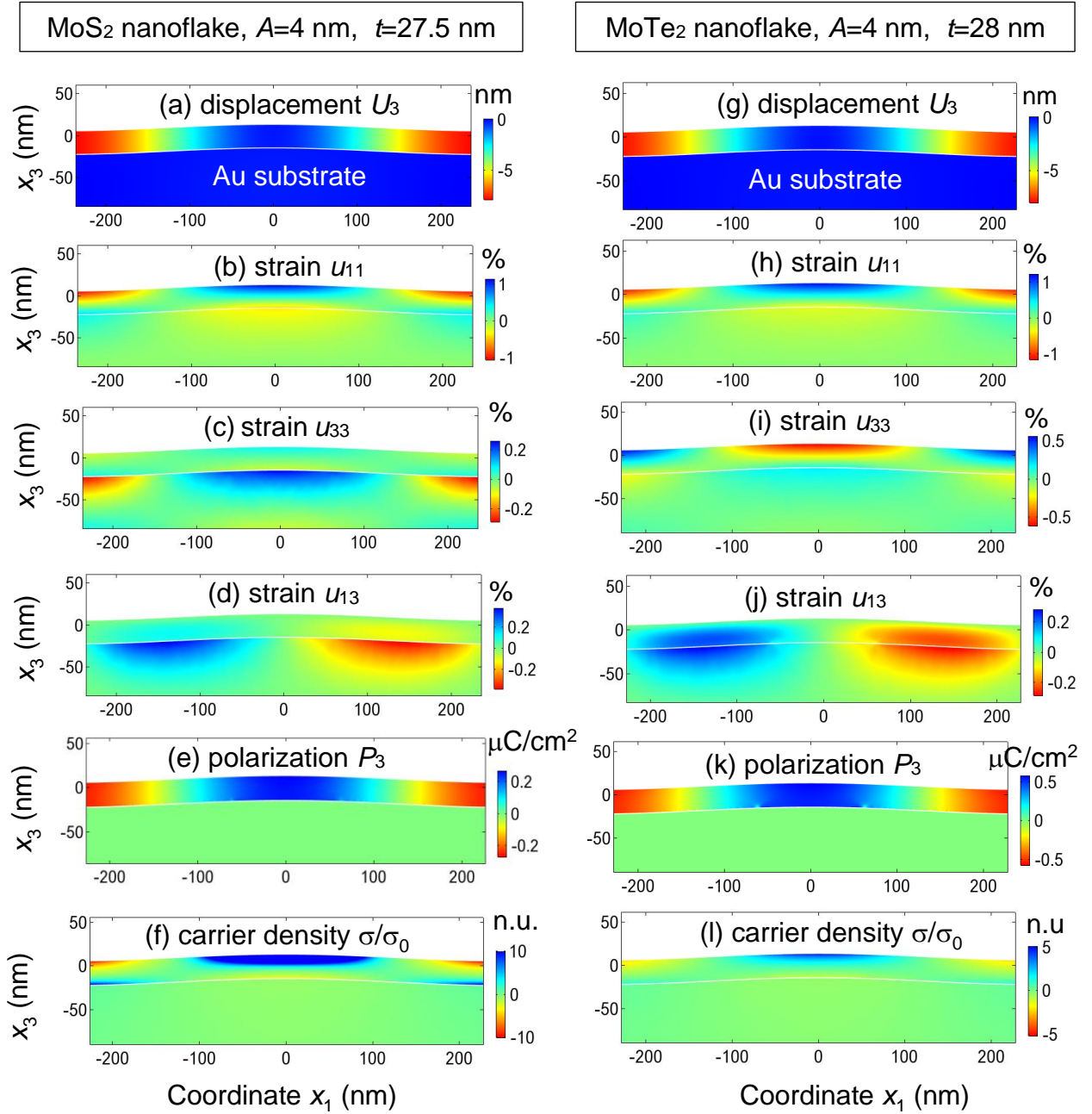

**FIGURE 3. Distributions polar and electronic properties in MoS$_2$ (a-f) and MoTe$_2$ (g-l) nanoflakes.** Bending-induced elastic displacement (**a, g**), strain tensor components (**b-d, h-j**), out-of-plane flexoelectric polarization (**e, k**), and relative static conductivity (**f, l**) in the cross-section of a MoS$_2$ (**a-f**) and MoTe$_2$ (**g-l**) nanoflake with thickness $t \approx 28$ nm, which is placed on a 200-nm thick rough substrate. The amplitude $A$ of the substrate corrugation is 4 nm, and the average period of the corrugation is $\lambda = 0.5$ μm. MoS$_2$, MoTe$_2$, and substrate parameters are listed in **Table I**.

The cross-sections of bending-induced elastic displacement, strain components, out-of-plane polarization for ultra-thin MoSTe nanoflakes (e.g. 10 layers or thinner, see **Figs. 4a-f**) looks only a bit different from thicker MX$_2$ nanoflakes (shown in **Figs. 3**) due to the thickness effect. When the MoSTe nanoflake thickness increases (see **Figs. 4a-f**), the overall picture looks as some "average" between the MoS$_2$ and MoTe$_2$ nanoflakes shown in **Figs. 3.** Note the possible appearance of the p-n junctions between the n-type and p-type regions in MXY nanoflakes, similarly to the case of MX$_2$ nanoflakes.

The influence of the surface-induced piezoelectric effect (present in 1H-MoSTe and absent for 1H and 2H-MoX$_2$) appears too small to affect significantly the polar and electronic properties distributions shown in **Fig. 4**. Hence even for the case of a JC MoSTe the dominant part of the polarization is the flexoelectric polarization. This conclusion is clearly seen from comparison of the scales in **Fig. S5a-b**, where the magnitude of the piezoelectric polarization is about 2 nC/cm$^2$, while the magnitude of the flexoelectric polarization is two orders higher – about 0.4 μC/cm$^2$ for the flake thickness ~ (4 – 20) nm**.** At that the contribution of the surface-induced piezoelectric effect decreases with the flake thickness according to Eq.(3) (compare brown and black curves in **Fig. S5a**).

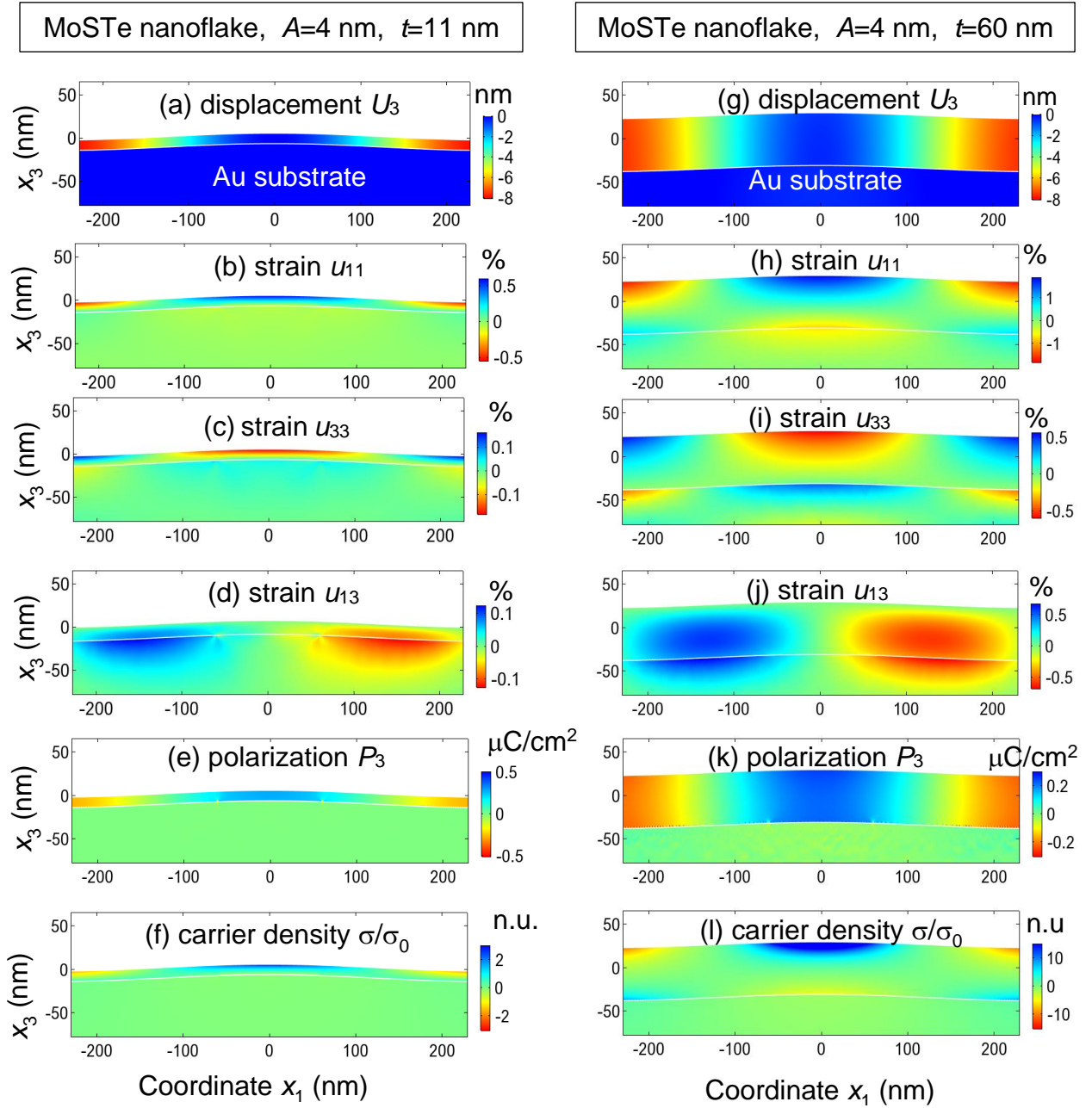

**FIGURE 4. Distributions polar and electronic properties in MoSTe nanoflakes with thickness 11 nm (a-f) and 60 nm (g-l).** Bending-induced elastic displacement **(a,g)**, strain tensor components **(b-d, h-j)**, out-of-plane flexoelectric polarization **(e,k)**, and relative static conductivity **(f,l)** in the cross-section of the nanoflake placed on a thick rough substrate. The amplitude $A$ of the substrate corrugation is 4 nm, and the average period of the corrugation is $\lambda = 0.5$ μm. MoSTe and substrate parameters are listed in **Table I.**

The curves in **Fig. 5** are calculated for different amplitude $A$ of substrate roughness $A = (0 - 11)$ nm and approximately the same thickness ~11 nm (corresponding to the range from 10 to 8 layers) of MoS$_2$, MoTe$_2$ and MoSTe nanoflakes. The bending-induced out-of-plane flexoelectric polarization and relative electron density monotonically increase with $A$ increase. At that the relative increase of the electron density for MoS$_2$ (>10$^3$ times) is much higher than that for a MoTe$_2$ (>10 times), and also than that for a MoSTe (>20 times), compare **Fig. 5b, 5d** and **5f**. The difference is related with significantly higher electronic deformation potential of MoS$_2$ (see **Table I**). The profiles of the out-of-plane flexoelectric polarization, which is proportional to the strain gradient, demonstrate visible deviations from the sinusoidal profile of substrate, since the top regions of the 10 curves plotted for $A = (2 - 11)$nm have two symmetric indentations, which height increases and location becomes more close to the peak $x_1 = 0$ with $A$ increase (see **Fig. 5a, 5c** and **5e**).

Bending dependence of polar and electronic properties of MoS$_2$ and MoTe$_2$ nanoflakes are shown in **Fig. 6a-b** and **6c-d**, respectively. The bending-induced elastic strain and strain gradient, out-of-plane flexoelectric polarization and relative electron density of the nanoflakes monotonically increase with the increase of corrugation amplitude $A$. The increase of the strain and electron density is linear at small $A < 5$ nm and becomes super-linear at $A > 10$ nm; it is much higher at the upper surface of the nanoflake in comparison with its bottom surface (compare red/brown and blue/black circles in **Fig. 6a** and **6d**). In contrast, the strain gradient and flexoelectric polarization, which increase linearly with the amplitude $A$ increase, are almost the same at the upper and bottom surfaces (compare red/brown and blue/black circles in **Fig. 6b** and **6c**). Note the qualitative similarity, but the quantitative difference between MoS$_2$ and MoTe$_2$ nanoflakes, which are the most pronounced for polarization (2 µC/cm$^2$ vs. 4 µC/cm$^2$), and relative electron density ($\gg$10$^3$ vs. 10$^2$) maximal values.

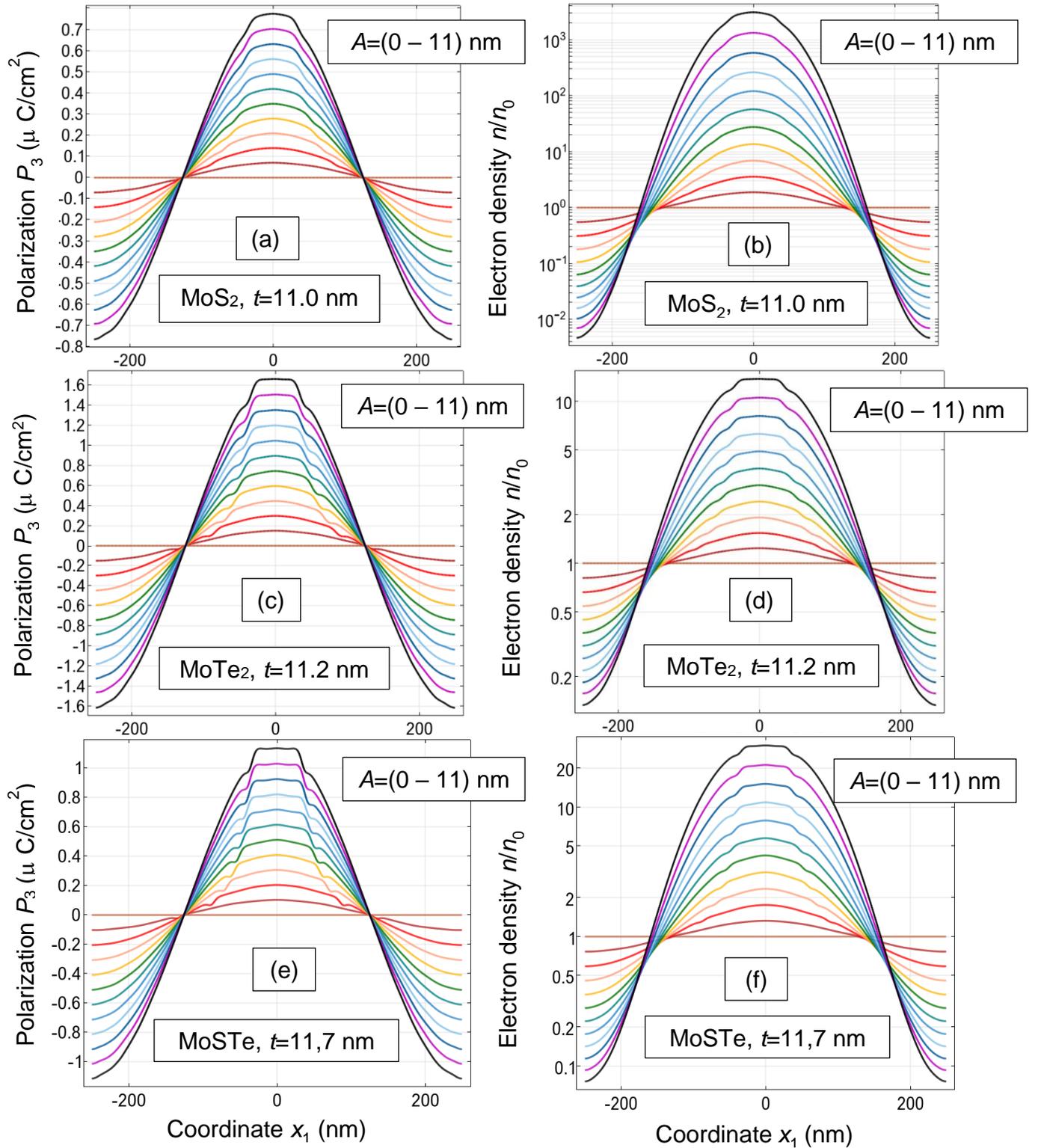

**FIGURE 5. Profiles of polar and electronic properties of MoS$_2$, MoTe$_2$ and MoSTe nanoflakes.** Bending-induced out-of-plane flexoelectric polarization **(a, c, e)**, and relative electron density **(b, d, f)** at the upper surface of the nanoflake with thickness $t \approx 11$ nm, which is placed on a thick rough substrate. Twelve curves (from "0"

brown line to the top black curve) corresponds to the different values of roughness amplitude $A = 0, 1, 2, 3, 4, \ldots 11$ nm. Nanoflakes and substrate parameters are listed in **Table I.**

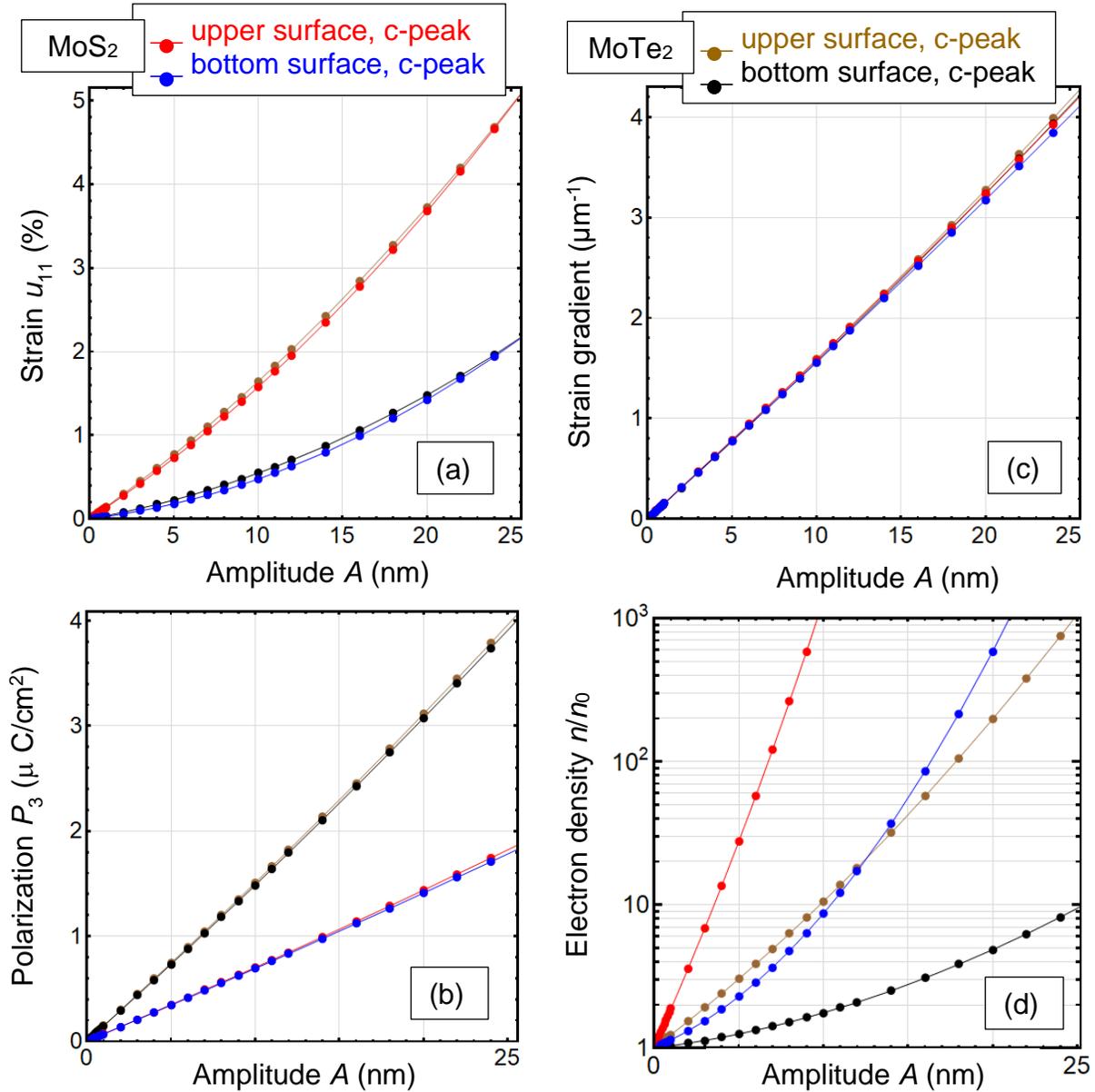

**FIGURE 6. Bending dependence of polar and electronic properties of MoS$_2$ (red and blue circles) and MoTe$_2$ (brown and black circles) nanoflakes.** The bending-induced elastic strain **(a)** and strain gradient **(b)**, out-of-plane flexoelectric polarization **(c)** and relative electron density **(d)** of a nanoflake via the corrugation amplitude $A$ of the rough substrate. The quantities **(a)-(d)** are shown at the upper (red and brown circles) and bottom (blue and black circles) surfaces of the nanoflake, which is placed at the peak of the substrate corrugation

(abbreviated as "**c-peak**", see **Fig. 2**). The flake thickness is 11 nm, the average period of corrugation $\lambda = 0.5$ μm. MoX$_2$ and substrate parameters are listed in **Table I.**

Thickness dependence of polar and electronic properties of MoS$_2$ and MoTe$_2$ nanoflakes are shown in **Fig. 7**. Blue/black and red/brown circles, which correspond to the corrugation peaks, where the strain and strain gradient are maximal, show the maximal increase of the flexoelectric polarization, and electron or hole densities. Green/purple circles, which correspond to the corrugation slope, where the strain and strain gradient are absent, show almost zero flexoelectric polarization and no increase in the electron or hole density. Most important that there is a pronounced maximum at the thickness dependences of the elastic strain, electron and hole conductivity of MoS$_2$ and MoTe$_2$ nanoflakes placed on a rough substrate with a sinusoidal corrugation profile. The optimal thickness is corrugation dependent, and it is 75 nm for MoS$_2$ and 80 nm MoTe$_2$ nanoflakes for the case of 4 nm corrugation height. The result opens the way for their geometry optimization.

Note, that corresponding bending and thickness dependence curves for MoSTe nanoflakes lay between MoS$_2$ and MoTe$_2$ nanoflakes, as anticipated from **Fig. 5**. They are not shown in **Figs. 6** and **7**, since the "mixture of curves" appears when we add two or three more curves.

Analysis of the above FEM results obtained for the different flake thickness (varying from 10 layers to 300 nm) and corrugation depth (varying from 0 to 25 nm), allows to corroborate the flexoelectric nature of the out-of-plane electric polarization [22] and establish the unambiguous correlation between the polarization and *n*-type (or *p*-type) static conductivity modulation caused by inhomogeneous elastic strains coupled with deformation potential, and strain gradients, which evolve in TMD nanoflake due to the adhesion between the flake surface and corrugated substrate.

These results can be useful for elaboration of nanoscale straintronic devices based on the MX$_2$ and MXY nanoflakes bended by a corrugated substrate. Principal schemes of diodes and bipolar transistors are presented in **Fig. 8**, where the bending profile of a semiconducting nanoflake controls the sharpness of p-n junctions between the regions with n-type (electron) and p-type (hole) conductivity**.** To create real prototypes of devices, one should take into account possible disadvantages related with a "doubled" geometry of the obtained p-n junctions. Namely we calculated more faint "extra" p-regions located at the flake-substrate boundary below the top n-regions, and "extra" n-regions located at the flake-substrate boundary below the top p-regions. However, for a specific device architecture the "doubled" geometry in complex with very high mechanical stability and flexibility of the nanoflakes-on-substrate may become a benefit.

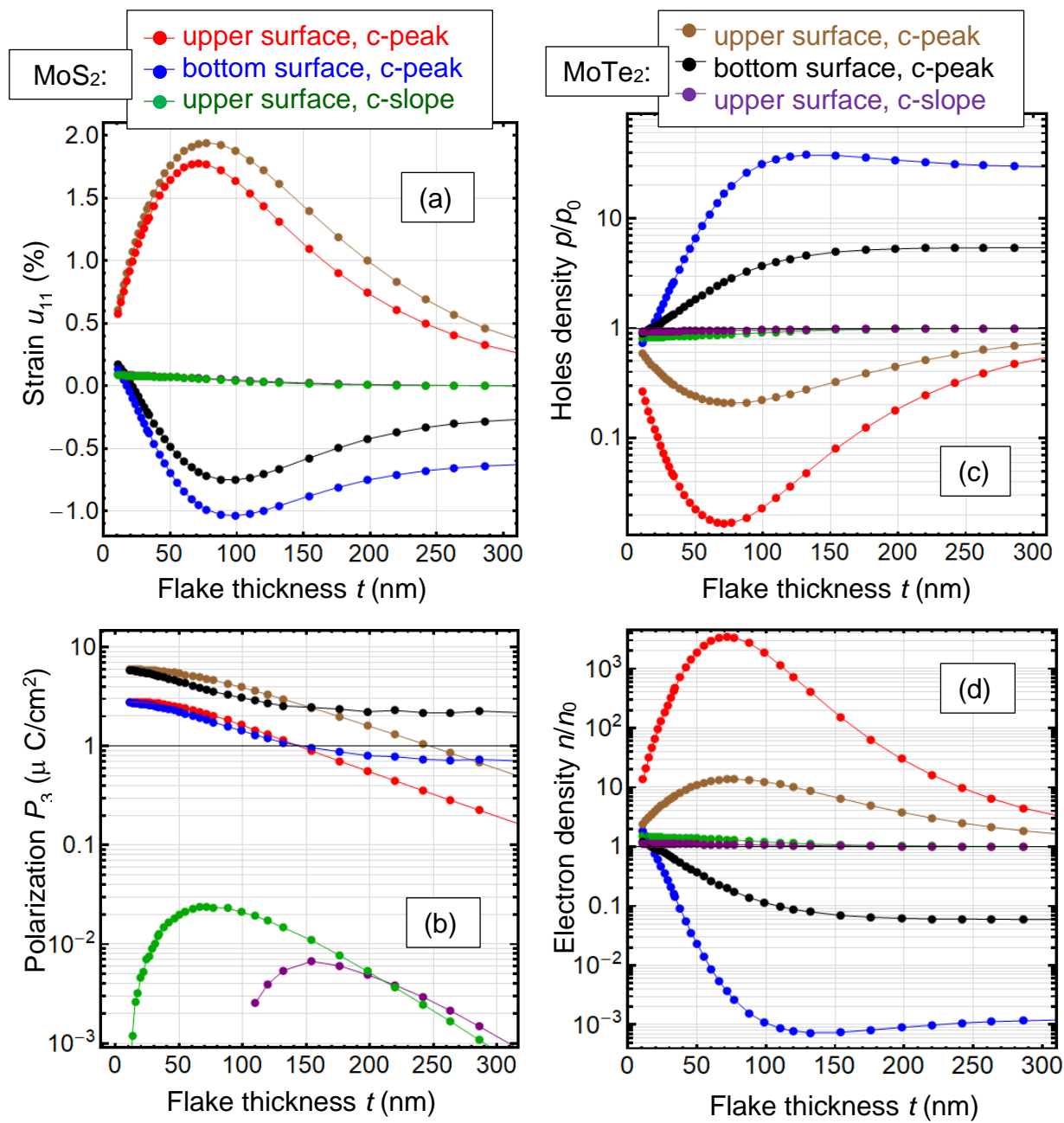

**FIGURE 7. Thickness dependence of polar and electronic properties of MoS$_2$ (red, green and blue circles) and MoTe$_2$ (brown, purple and black circles) nanoflakes.** The bending-induced elastic strain **(a)**, out-of-plane flexoelectric polarization **(b)**, hole **(c)**, and **(d)** electron densities via the thickness of a nanoflake placed on a rough substrate. The quantities **(a)-(d)** are shown at the upper (red and brown circles) and bottom (blue and black circles) surfaces of the nanoflake, which is placed at the peak of the substrate corrugation (abbreviated as "**c-peak**", see

**Fig. 2**). Green and purple circles correspond to the corrugation slope (abbreviated as "**c-slope**", see **Fig. 2**). The amplitude *A* of substrate roughness is 4 nm, the average period of its corrugation $\lambda = 0.5$ μm. MoX$_2$ and substrate parameters are listed in **Table I.**

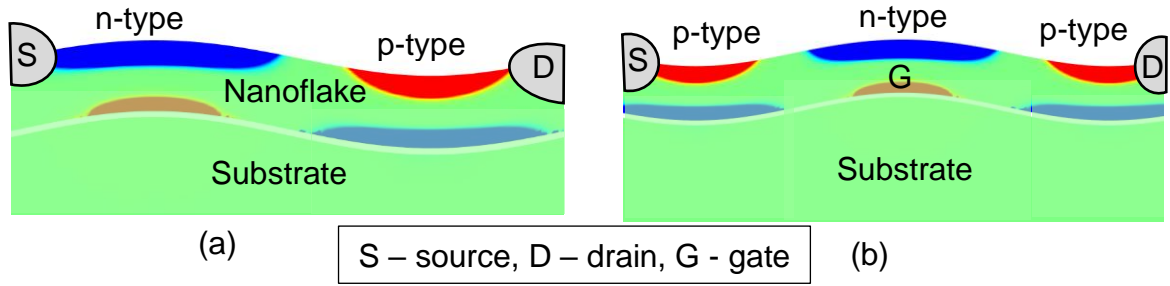

**FIGURE 8.** Principal schemes of a diode **(a)** and bipolar transistor **(b)** based on a 13 nm thick MoSTe nanoflakes bended by a corrugated substrate. The regions with n-type (electron) and p-type (hole) conductivity are shown.

To verify theoretical results, the structure of MoS$_2$ nanoflakes was examined by a scanning tunneling microscopy (**STM**) in non-vacuum conditions (see details in **Appendix C** [49]). As substrates we used highly-ordered pyrolytic graphite (HOPG) and Au (111) on mica. STM images, shown in **Fig. S6 – S7**, revealed MoS$_2$ nanoflakes of different shape and size. Local volt-ampere characteristics (I-V curves) measured at the several close points of a nanoflake are also shown. The significant variability of the I–V curves and typical jumps on them indicate the randomness of a tunneling current path associated with the strong local inhomogeneity of electroconductivity near the surface of MoS$_2$ nanoflakes, which is most likely caused by the shape changes of the nanoflakes and by the evolution of local electric fields under the current flow. In fact, the strong local inhomogeneity of the MoS$_2$ nanoflake conductivity can be determined by the bending of its surface in accordance with FEM results presented in this work. Hence STM results indirectly indicates a significant dependence of free carrier concentration in MoS$_2$ nanoflakes on their bending and strain gradient. The quantitative description of these results requires theoretical calculations of local electric fields coupled to elastic strains for different substates and nanoflake shape/sizes. This task will be performed in near future.

## V. SUMMARY

Using FEM, we calculated the elastic and electric fields, flexoelectric polarization and free charge density for a TMD nanoflake placed on a rough substrate with a sinusoidal profile of the corrugation principal component. Analysis of FEM results obtained for the different flake thickness (varying from 10 layers to 300 nm) and corrugation depth (varying from 0 to 25 nm), allows to corroborate the flexoelectric nature of the out-of-plane electric polarization and establish the unambiguous correlation between the polarization and static conductivity modulation caused by inhomogeneous elastic strains coupled with deformation potential, and strain gradients, which evolve in the TMD nanoflake due to the adhesion between the flake surface and corrugated substrate.

We revealed a pronounced maximum at the thickness dependences of the electron and hole conductivity of $MoX_2$ and MXY nanoflakes placed on a substrate with a sinusoidal corrugation. Namely the conductivity is maximal for a (75 − 80) nm thick flakes placed on a rough substrate with the 4 nm corrugation height. This result opens the way for the nanoflakes geometry optimization towards significant improvement their polar and electronic properties, necessary for their advanced applications in nanoelectronics and memory devices. Specifically, obtained results can be useful for elaboration of nanoscale straintronic devices based on the bended $MX_2$ and MXY nanoflakes, such as diodes and bipolar transistors with a bending-controllable width of p-n junctions.

**Acknowledgements.** This effort (S.V.K.) is based upon work supported by the US Department of Energy (DOE), Office of Science, Basic Energy Sciences (BES), Materials Sciences and Engineering Division and was performed at the Oak Ridge National Laboratory's Center for Nanophase Materials Sciences (CNMS), a US Department of Energy, Office of Science User Facility. This work (Y.K.) was supported by the National Research Foundation of Korea (NRF) grant funded by the Korea government (MSIT) (No. 2020R1F1A1072355). A.N.M, H.V.S., Y.Y.L. and G.I.D. work has received funding from the National Research Foundation of Ukraine (Grant application 2020.02/0027, Contrast N 52/02.2020).

**Authors' contribution.** A.N.M., Y.K. and S.V.K. generated the research idea. A.N.M evolved the theoretical model, state the problem, interpreted numerical results obtained by E.A.E. and H.V.S., and jointly with Y.K. and S.V.K. wrote the manuscript draft. Y.Y.L. performed STM measurements. G.I.D., Y.K. and S.V.K. worked on the results discussion and manuscript improvement.

# Supplementary Materials to
# "Correlation between corrugation-induced flexoelectric polarization and conductivity of low-dimensional transition metal dichalcogenides"

Anna N. Morozovska[1*], Eugene A. Eliseev[2], Hanna V. Shevliakova[1,3], Yaroslava Yu. Lopatina[1], Galina I. Dovbeshko[1], Maya D. Glinchuk[2], Yunseok Kim[4†], and Sergei V. Kalinin[5,‡]

[1]*Institute of Physics, National Academy of Sciences of Ukraine,*
*46, pr. Nauky, 03028 Kyiv, Ukraine*

[2] *Institute for Problems of Materials Science, National Academy of Sciences of Ukraine,*
*Krjijanovskogo 3, 03142 Kyiv, Ukraine*

[3] *Department of Microelectronics, National Technical University of Ukraine "Igor Sikorsky Kyiv Polytechnic Institute", Kyiv, Ukraine*

[4] *School of Advanced Materials Science and Engineering, Sungkyunkwan University (SKKU), Suwon 16419, Republic of Korea*

[5] *The Center for Nanophase Materials Sciences, Oak Ridge National Laboratory,*
*Oak Ridge, TN 37922*

---

[*] Corresponding author 1: anna.n.morozovska@gmail.com

[†] Corresponding author 2: yunseokkim@skku.edu

[‡] Corresponding author 3: sergei2@ornl.gov

**APPENDIX A. An Estimate of the Interaction Force between a SL-MX$_2$ and Substrate**

In a general case, different types of vdW interactions, such as attractive potentials for dipole–dipole, dipole–induced dipole, transient induced dipole–induced dipole and dispersive (London) interactions between several molecules or atomic groups, contribute to the "effective" interfacial cohesive energy $J_{EF}$, which value we can vary in the range from 0.001 J/m$^2$ to 1 J/m$^2$, remaining significantly smaller than the MX$_2$ bond energy density [1]. As the first step, let us estimate the interaction force between the first layer of chalcogen atoms and substrate, based on the assumption that van der Waals (**vdW**) interaction can determine it. We assume that the separation of thickness $h$ less than 0.3 nm is present between the first layer of MX$_2$ (or MXY) and substrate [2]. The net vdW force (per unit area, in J/m$^3$) between two flat plates is [3]:

$$F_{VdW} \approx -\frac{A_H}{6\pi h^3}, \tag{A.1a}$$

where the typical values for Hamaker constant $A_H$ of different materials are in the order of 10$^{-19}$ J. The constant $A_H$ depends strongly on the contact conditions. Assuming a constant separation between two surfaces, $h$, the specific interfacial interaction energy, $W_{VdW}$, per unit area is given by

$$W_{VdW} \approx +\frac{A_H}{12\pi h^2}, \tag{A.1b}$$

Typical value of the interfacial energy $W_{VdW}$ due to vdW interaction is less than 0.01 J/m$^2$ and the "effective" constant $A_H^{EF}$ can vary from 10$^{-20}$ J to 10$^{-17}$ J [3]. The "effective" vdW force density can decay with the distance as $S_A/h$ and for sufficiently small separations they may easily become the dominant force.

The strong power decrease of the interaction force allows us to assume that the amplitude of atomic displacement, schematically shown in **Fig.1c-d** and **1g-h**, is the smallest for the first layer of X-atoms, then significantly increase the for the middle M-atoms, and very strongly increase for the second layers of X- (or Y) atoms in comparison with the first layer of X-atoms. The displacement amplitude of the first X-layer depends on the substrate material and interfacial conditions. It can be very small if effective interaction constant $A_H^{EF}$ is high (e.g. close to maximal value ~10$^{-17}$ J), and comparable with interatomic distance if $A_H^{EF}$ is small (e.g. close to minimal value ~10$^{-20}$ J). The case of high $A_H^{EF}$, that corresponds to the atomic "clamping" by a rigid substrate, is schematically shown in **Fig.1c** and **1g.**

# APPENDIX B. Supplementary Figures

Correlation between elastic strain, its gradient, polar and electronic properties of TMDs nanoflakes are shown in **Figs. S1-S5.** Bending-induced elastic displacement, strain, strain gradient, out-of-plane electric polarization, and electron density are shown at the surface of a nanoflake of length $L$, width $W$ and thickness $t$, at that $L \gg W \gg t$. MoS$_2$ and MoTe$_2$ nanoflakes are placed on a rough gold substrate with a corrugation profile $\vec{x}_s = \left\{A \cos\left(\frac{2\pi}{\lambda}x_1\right), 0, 0\right\}$, and the flake displacement, strain, strain gradient, out-of-plane electric polarization are modulated by the corrugation profile, being maximal at the corrugation tops and minimal in the valleys.

Different curves in **Fig. S1** and **S3** are calculated for different thickness $t = (11 - 27)$ nm of MoS$_2$ and MoTe$_2$ nanoflakes and the same amplitude $A$ of substrate roughness $A = 4$ nm. As one can see the bending-induced elastic displacement and relative electron density increase with $t$ increase (see **Fig. S1a, S1d** and **S3a, S3d**). At that the relative increase of the electron density for MoS$_2$ (100 times) is much higher than that for a MoTe$_2$ (5 times), compare **Fig. S1d** and **S3d**. The difference is related with significantly higher electronic deformation potential of MoS$_2$, $\Sigma_e = -11.14$ eV (see **Table I**). The strain gradient and the out-of-plane electric polarization look almost thickness-independent, but contain visible deviations from the sinusoidal profile of substrate, since the top regions of all 4 curves have two symmetric indentations at $x_1 \approx \pm 50$ nm (see **Fig. S1b, S1c** and **S3b, S3c**).

Different curves in **Fig. S2** and **S4** are calculated for different amplitude $A$ of substrate roughness $A = (0 - 11)$ nm and approximately the same thickness ~11 nm (e.g. 10 and 8 layers) of MoS$_2$ and MoTe$_2$ nanoflakes. As one can see the bending-induced elastic displacement, strain, out-of-plane electric polarization, and relative electron density monotonically increase with $A$ increase. At that the relative increase of the electron density for MoS$_2$ ($>10^3$ times) is much higher than that for a MoTe$_2$ ($>10$ times), compare **Fig. S2d** and **S4d**. The difference is related with significantly higher electronic deformation potential of MoS$_2$ (see **Table I**). The profiles of the out-of-plane displacement are smooth and follow the sinusoidal profile of the corrugated substrate (see **Fig. S2a** and **S4a**). The profiles of strain, and out-of-plane electric polarization (~to the strain gradient) demonstrate visible deviations from the sinusoidal profile of substrate, since the top regions of all 10 curves plotted for $A = (2 - 11)$ nm have two symmetric indentations, which height increases and location becomes more close to the top $x_1 = 0$ with $A$ increase (see **Fig. S2b, S2c** and **S4b, S4c**).

For the case of CJ 1H-MoSTe the dominant part of the polarization is the flexoelectric polarization. This conclusion is clearly seen from comparison of the scales in **Fig. S5a-b**, where the magnitude of the piezoelectric polarization is about 2 nC/cm$^2$, while the magnitude of the flexoelectric polarization is two orders higher – about 0.4 μC/cm$^2$ for the flake thickness ~ (4 – 20) nm. The total polarization apparently coincides with the flexoelectric polarization. At that the contribution of the surface-induced piezoelectric effect decreases with the flake thickness (compare brown and black curves in **Fig. S5a**).

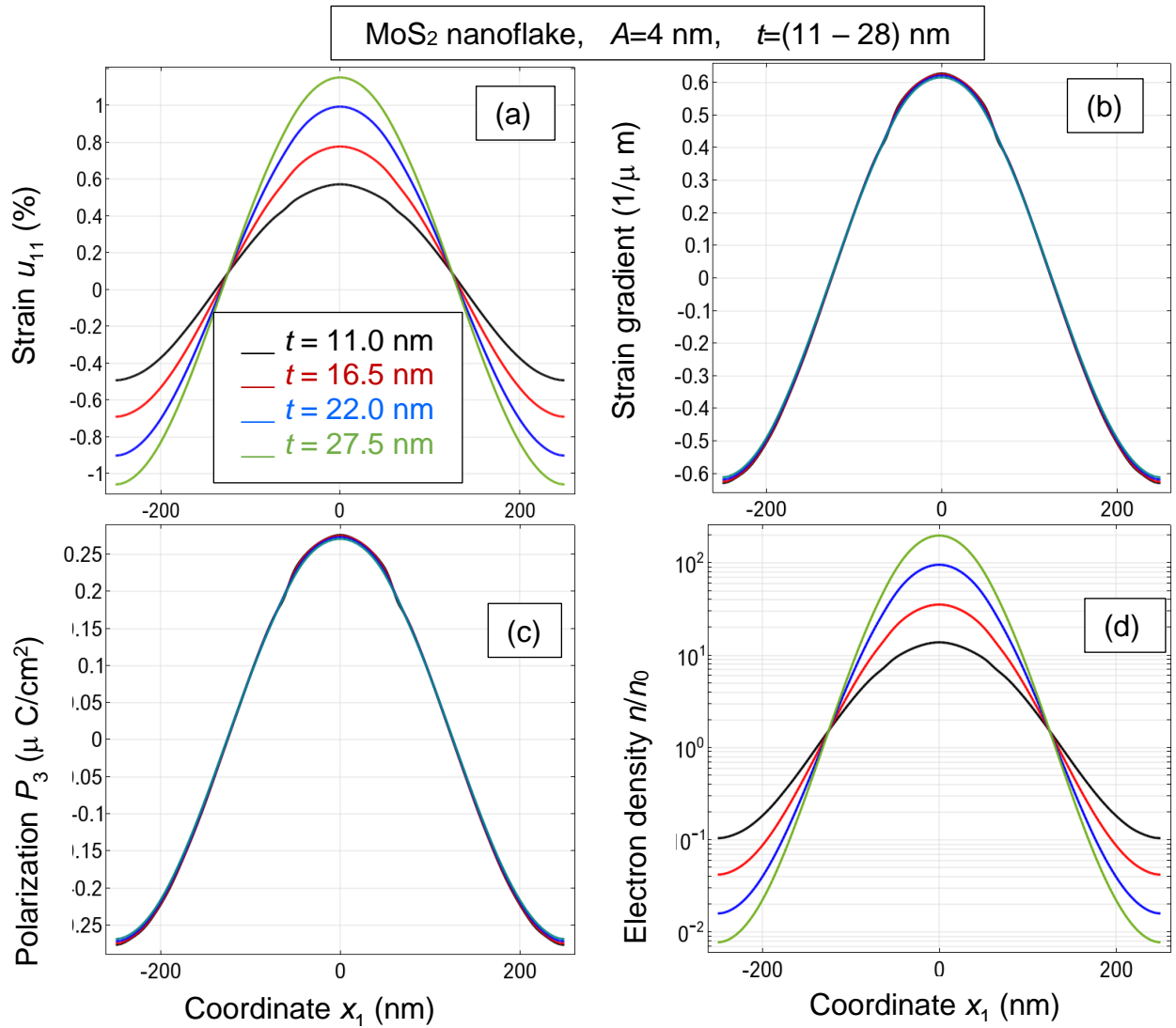

**FIGURE S1. Correlation between polar and electronic properties of MoS$_2$ nanoflakes.** Bending-induced elastic strain (**a**), strain gradient (**b**), out-of-plane electric polarization (**c**), and relative static electron conductivity

**(d)** at the top surface of a MoS$_2$ nanoflake with different thickness $t$ = 11 nm (black curves), 16.5 nm (red curves), 22 nm (blue curves) and 27.5 nm (green curves) on a corrugated Au-substrate. The amplitude $A$ of Au-substrate roughness is 4 nm, the average period of corrugation $\lambda$ =0.5 μm. Adhesive stiffness is 20 GPa, $\delta h_0 = 1.1$ nm. MoS$_2$ parameters are listed in **Table I.**

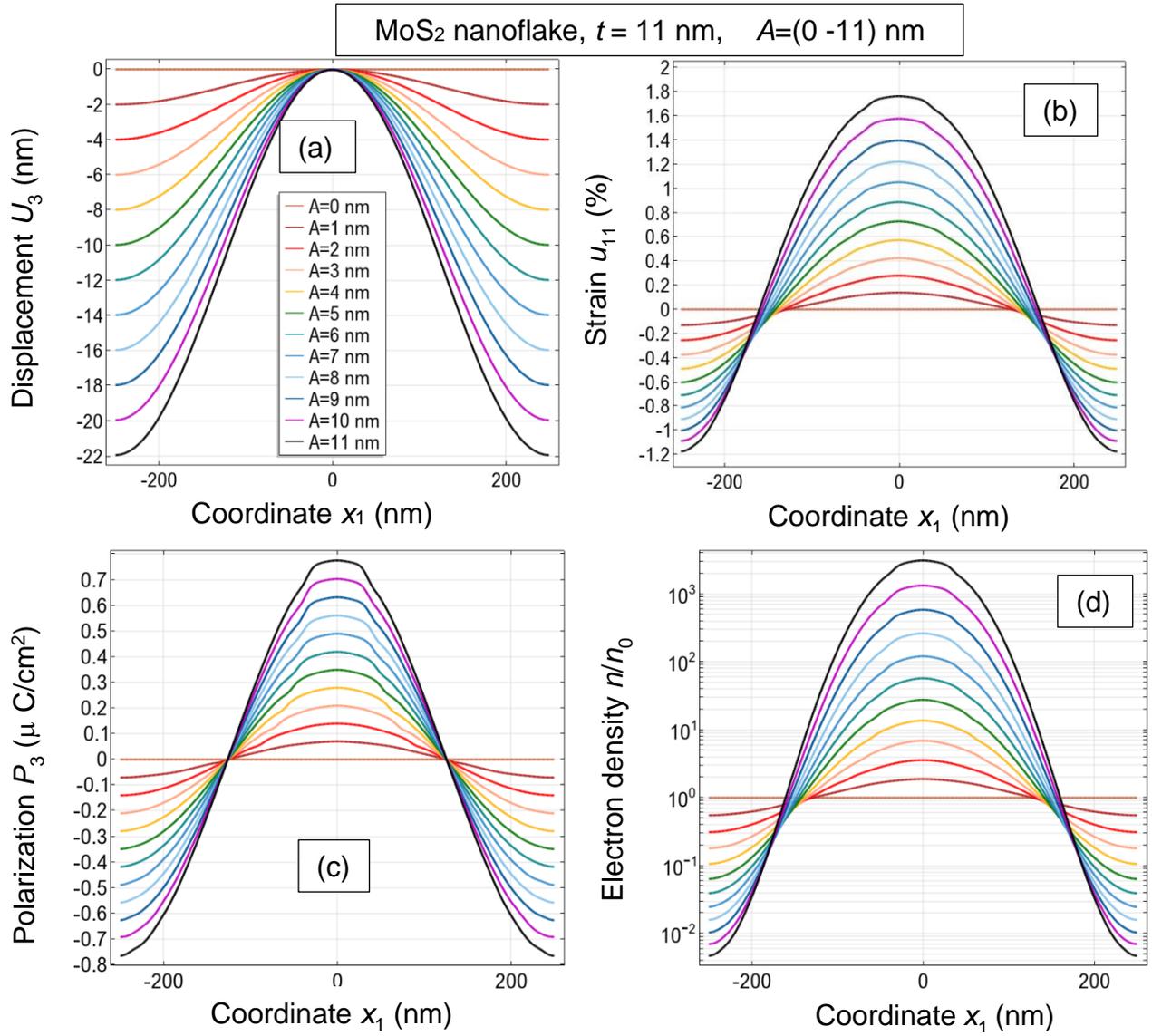

**FIGURE S2. Correlation between polar and electronic properties of MoS$_2$ nanoflakes.** Bending-induced elastic displacement **(a)**, strain **(b)**, out-of-plane electric polarization **(c)**, and relative electron density (i.e. static conductivity) **(d)** at the top surface of MoS$_2$ nanoflake with thickness $t$ =11 nm on the Au-substrate with different values of substrate roughness amplitude $A$=0, 1, 2, 3, 4,…11 nm (see label in plot (a)), the average period of corrugation $\lambda$ =0.5 μm. Adhesive stiffness is 20 GPa, $\delta h_0 = 1.1$ nm. MoS$_2$ parameters are listed in **Table I.**

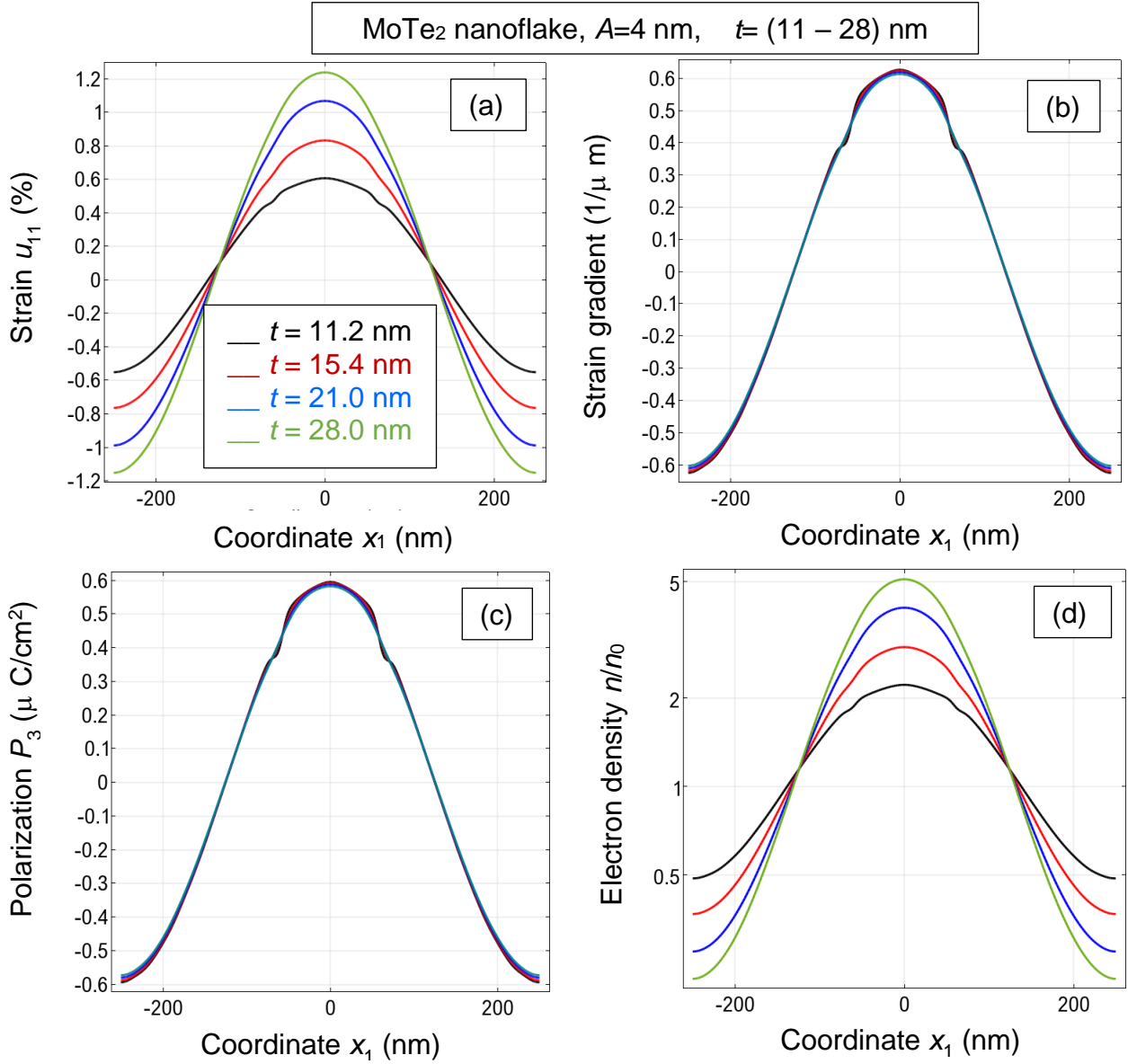

**FIGURE S3. Correlation between polar and electronic properties of MoTe$_2$ nanoflakes.** Bending-induced elastic strain **(a)**, strain gradient **(b)**, out-of-plane electric polarization **(c)**, and relative static electron conductivity **(d)** at the top surface of a MoTe$_2$ nanoflake with different thickness $t = 11$ nm (black curves), 16 nm (red curves), 22 nm (blue curves) and 27 nm (green curves) on the corrugated Au-substrate. The amplitude $A$ of the substrate roughness is 4 nm, the average period of corrugation $\lambda = 0.5$ μm. Adhesive stiffness is 20 GPa, $\delta h_0 = 1.1$ nm. MoTe$_2$ parameters are listed in **Table I.**

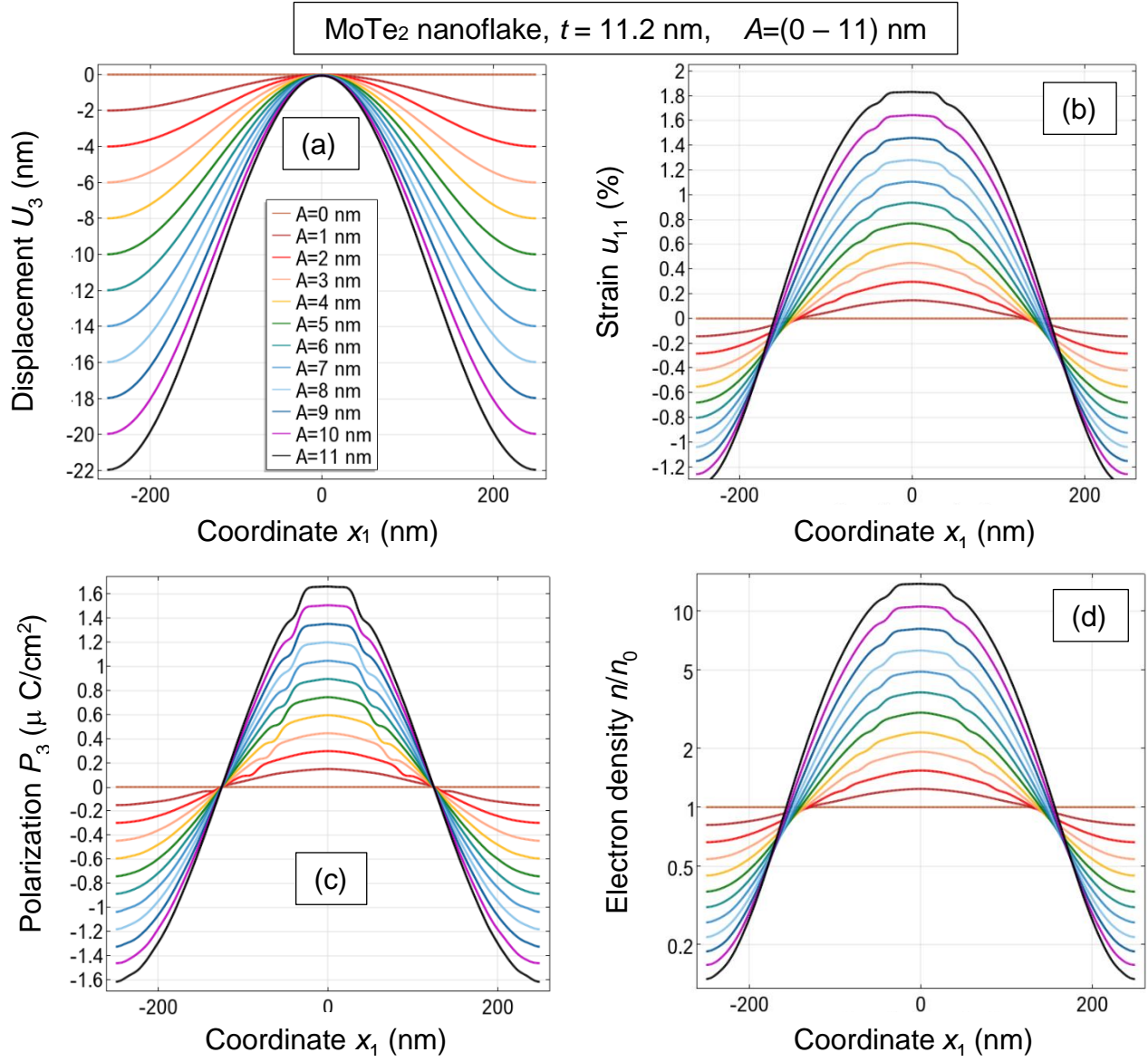

**FIGURE S4. Correlation between polar and electronic properties of MoTe$_2$ nanoflakes.** Bending-induced elastic displacement **(a)**, strain **(b)**, out-of-plane electric polarization **(c)**, and relative electron density (i.e. static conductivity) **(d)** at the top surface of MoTe$_2$ nanoflake with thickness $t$ =11 nm on the Au-substrate with different values of substrate roughness amplitude $A = 0, 1, 2, 3, 4, ... 11$ nm (see label in plot (a)), the average period of corrugation $\lambda$ =0.5 μm. Adhesive stiffness is 20 GPa, $\delta h_0 = 1.1$ nm. MoTe$_2$ parameters are listed in **Table I.**

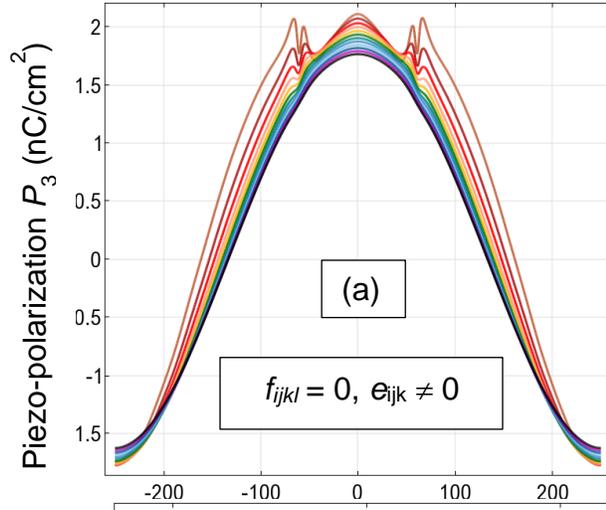
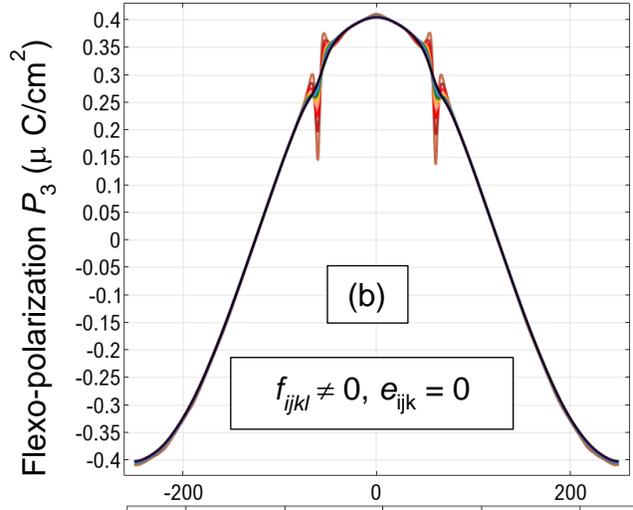
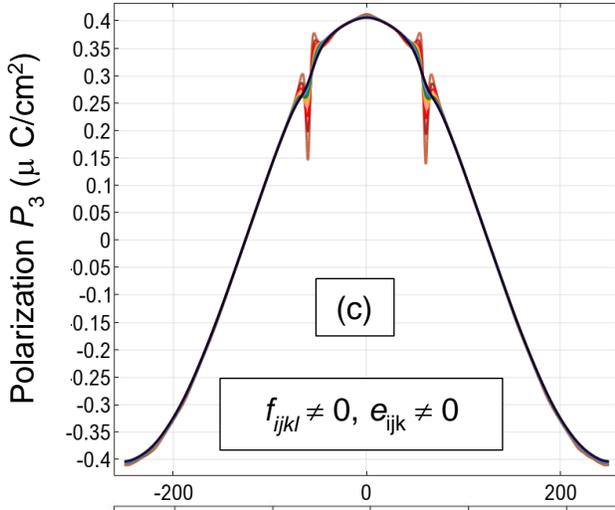
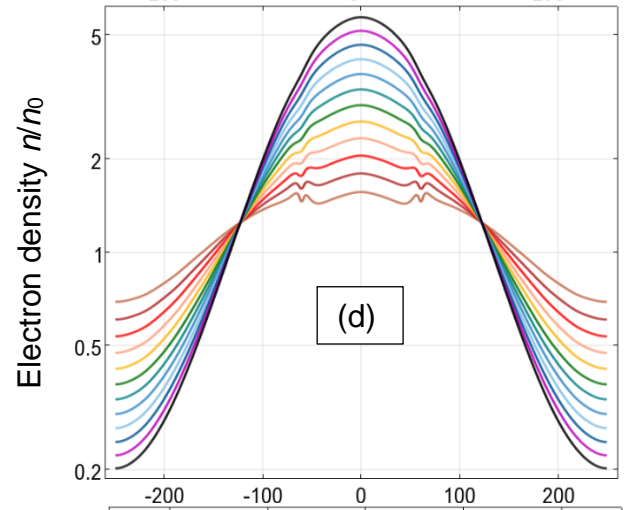
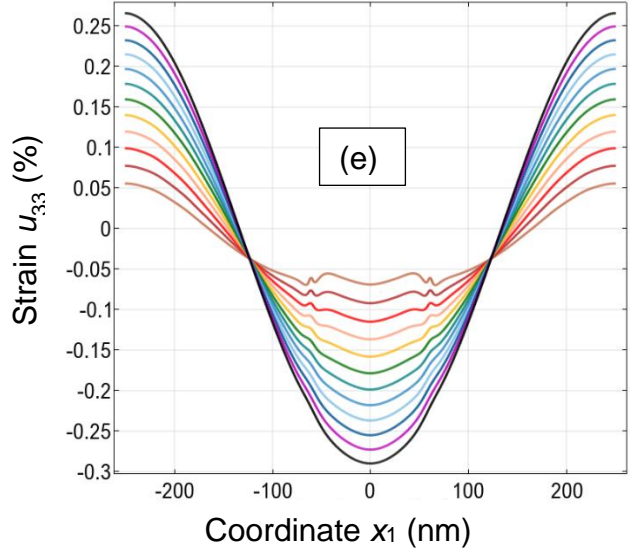
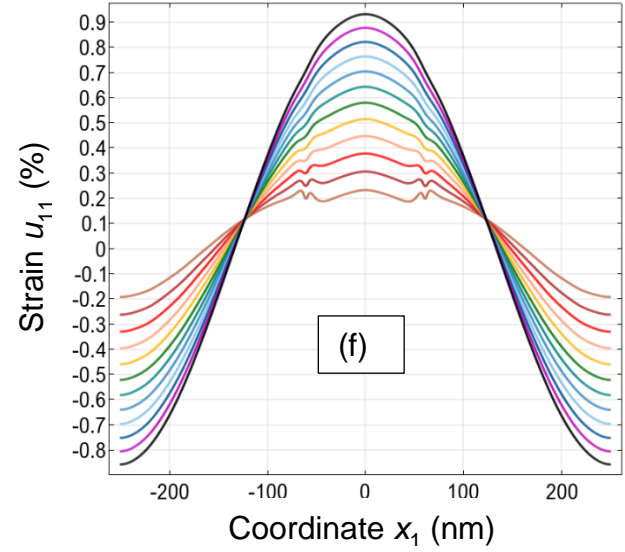

**FIGURE S5. Correlation between polar and electronic properties of MoSTe nanoflakes.** Piezo-induced **(a)**, bending-induced **(b)**, and total out-of-plane electric polarization **(c)**, and relative static electron conductivity **(d)** at the top surface of a nanoflake with different thickness *t* from 3.9 nm (brown curves) to 18.2 nm (black curves) with a 1.3 nm step. The flake is placed on a corrugated Au-substrate. The amplitude *A* of Au-substrate roughness is 4 nm, the average period of corrugation $\lambda$ =0.5 μm. Adhesive stiffness is 20 GPa, $\delta h_0 = 1.1$ nm. MoSTe parameters are listed in **Table I.**

## APPENDIX C. Scanning tunneling microscopy (STM) of MoS2 nanoflakes

The structure of $MoS_2$ nanoflakes was determined by STM in non-vacuum conditions. The STM method was used to detect the adsorbed $MoS_2$ nanoflakes, deposited on the surface from a water solution. Droplets with nanoflakes dispersed in water were deposited on a pristine clean substrate. The water was evaporated after the deposition. As substrates we used high-ordered pyrolytic graphite (HOPG) and Au (111) on mica. The probes were prepared by the mechanical sharpening of Pt-Ir (80:20) rod with a 0,25 mm diameter. Typical scanning parameters are voltage $U_t$ = (0.5-1.8) V and current $I_t$ = 70-200 pA. The scanning rate varies from 2 to 3μm/s. STM images were obtained in the constant current mode. The proposed deposition technology allows STM to resolve separate scattered nanoflakes and their segregates.

STM images revealed $MoS_2$ nanoflakes of different shape and size (see **Figs. S6 – S7**). The topography of the surface (Z-coordinate) along the direction AB is also shown in the figures. The diffuse boundaries of the nanoflakes indicate a doubling of the STM tip, but this deficiency does not affect the qualitative analysis. Allowing for the tip doubling effect, the following parameters of nanoflakes were determined: lateral dimensions ranged from 150 to 800 nm, thickness from 10 to 30 nm. We assume that the objects thicker than 40 nm are agglomerates of nanoflakes. Zoomed-scale images revealed the strong inhomogeneity of STM contrast on the surface of $MoS_2$ nanoflakes, which should be taken into account by scanning tunneling spectroscopy (**STS**).

Three local volt-ampere characteristics (I–V curves), measured at the points marked by marked by a circle with a central yellow-red rectangle, are also shown in **Figs. S6 – S7**. The significant variability of the I–V curves and typical jumps on them indicate the randomness of a tunneling current particular path associated with the strong local inhomogeneity of electrical conductivity near the surface of $MoS_2$ nanoflakes, which discrepancy is most likely caused by the changes in the shape of the nanoflakes and by the evolution of local electric field under the current flow.

In fact, the strong local inhomogeneity of the near-surface electrical conductivity of $MoS_2$ nanoflakes can be caused by the deformation of the $MoS_2$ surface according to the FEM results presented in this work. The strong local inhomogeneity indirectly indicates a significant dependence of the free carrier concentration in $MoS_2$ nanoflakes on their deformation and strain gradient.

The charge inhomogeneity of electrical conductivity is bipolar, as can be seen from different I–V curves. The critical voltage at which the tunnel current begins to rise sharply (corresponding to the opening of the I–V curves) depends on the substrate. It is much higher for HOPG substrate compared to Au (111) substrate (compare I–V curves in **Figs. S6** and **S7**). The quantitative description of these results requires theoretical calculations of local electric fields coupled with bending and elastic strains gradients for different substates and nanoflake shape/sizes. This task will be performed in near future.

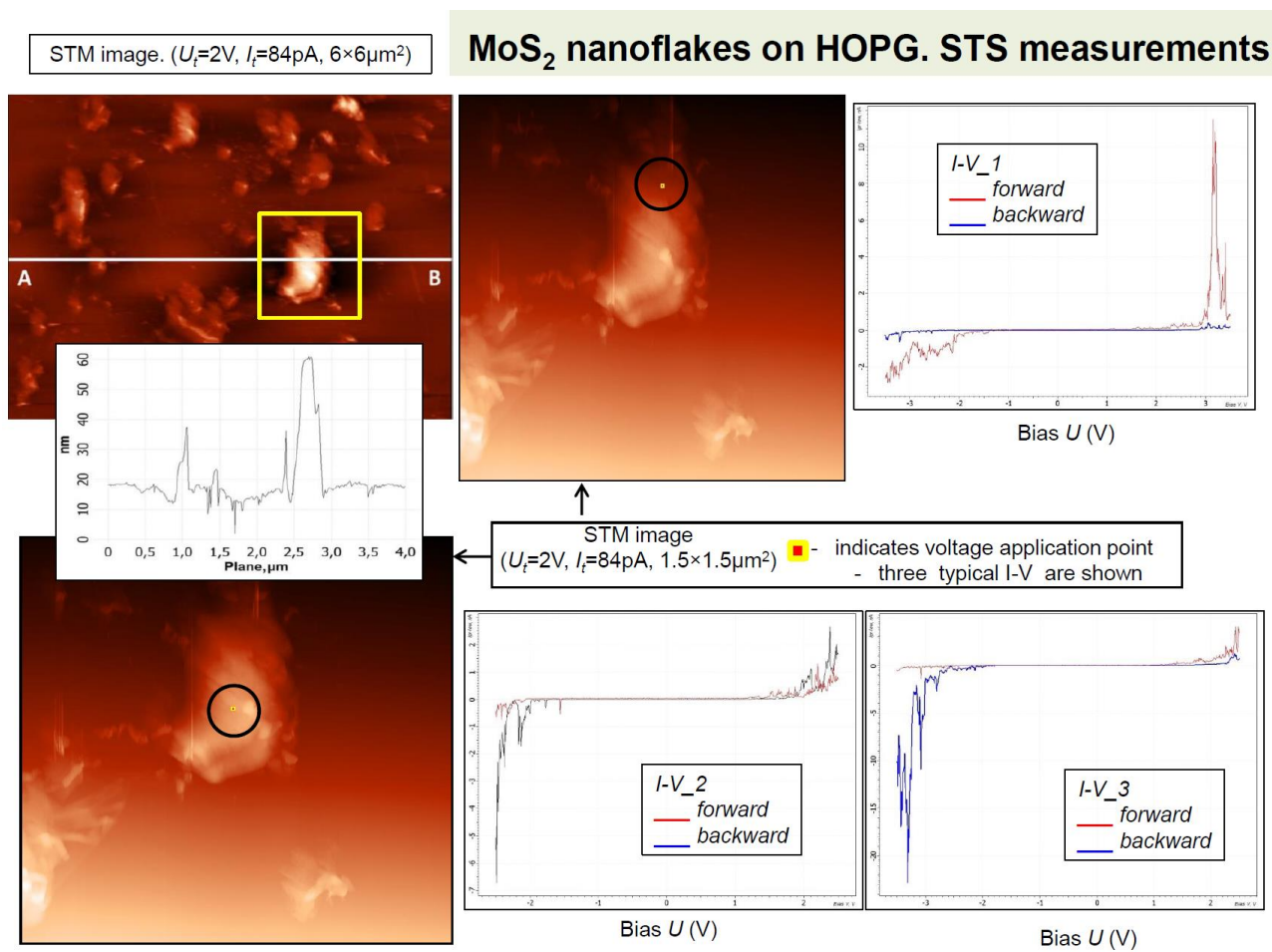

**FIGURE S6.** STM image of $MoS_2$ nanoflakes on HOPG ($U_t$ = 2 V, $I_t$ = 82 pA, scanned area 4 × 4 μm$^2$). The topography of the surface (Z-coordinate) along the direction AB and three local volt-ampere characteristics measured at the points marked by a circle with a central yellow-red rectangle.

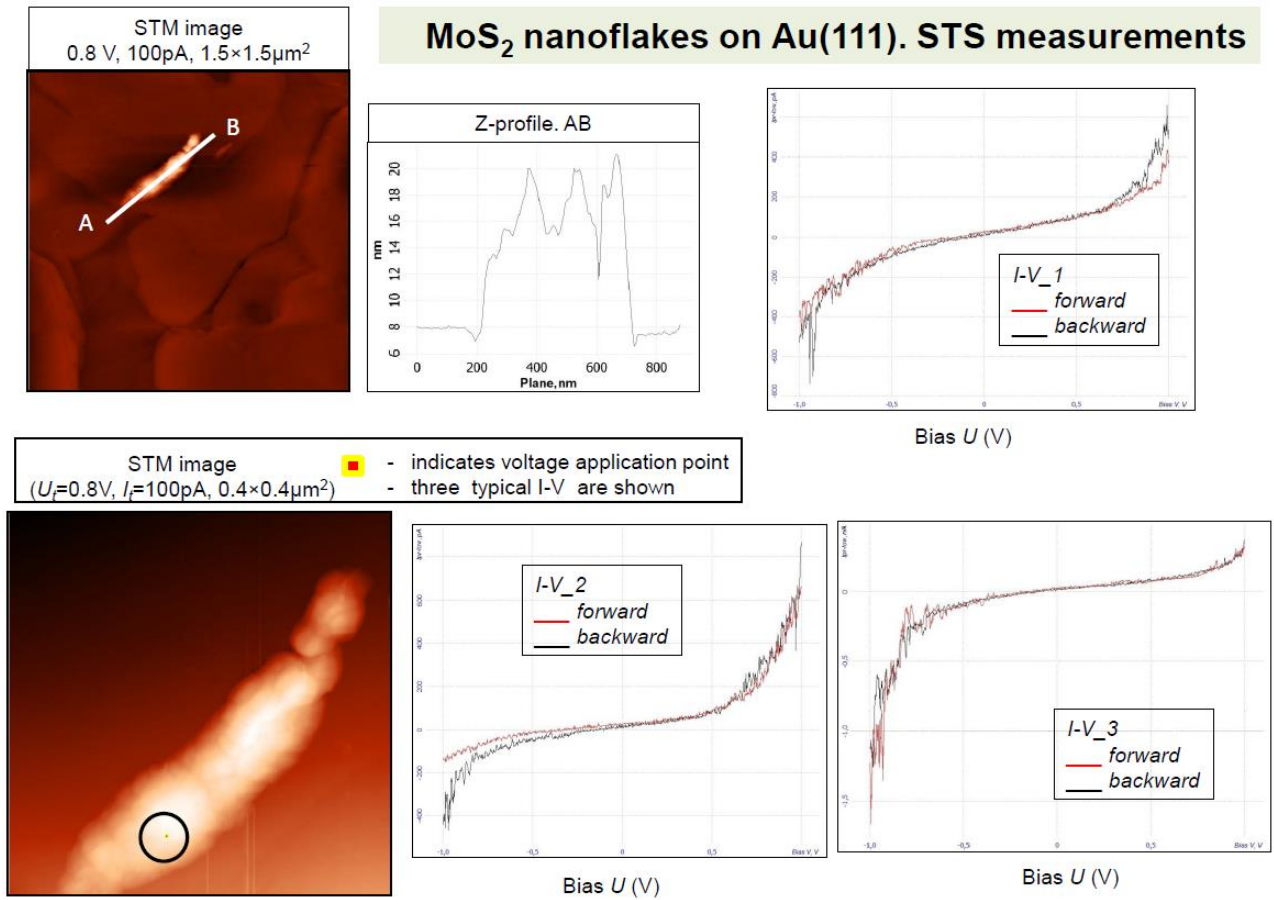

**FIGURE S7.** STM image of MoS$_2$ nanoflakes on Au(111) ($U_t$ = 2 V, $I_t$ = 82 pA, scanned area 4 × 4 μm$^2$). The topography of the surface (Z-coordinate) along the direction AB and three local volt-ampere characteristics measured at the points marked by a circle with a central yellow-red rectangle.

[3] T.F. Kuech, Integration of Dissimilar Materials, *in* Comprehensive Semiconductor Science and Technology, vol. 4, edited by P. Bhattacharya, R. Fornari and H. Kamimura (Elsevier Science, 2011). https://www.sciencedirect.com/topics/chemistry/van-der-waals-force